\documentclass[12pt,journal,onecolumn]{IEEEtran}

\usepackage{cite,authblk,rotating,setspace,amsmath,amssymb,theorem,epstopdf}
\usepackage[caption=false, font=footnotesize]{subfig}

\newtheorem{theorem}{Theorem}
\newtheorem{lemma}{Lemma}
\newtheorem{corollary}{Corollary}
\newtheorem{definition}{Definition}
\newenvironment{Proof}[1]{\medskip\par\noindent{\bf Proof:\,}\,#1}{{\mbox{\,$\blacksquare$}\par}}

\IEEEoverridecommandlockouts
\allowdisplaybreaks

\begin{document}

        \title{Age of Information in G/G/1/1 Systems: \\Age Expressions, Bounds, Special Cases, and Optimization}

        \author[1]{Alkan Soysal}
        \author[2]{Sennur Ulukus}
        \affil[1]{\normalsize Department of Electrical and Electronics Engineering,
Bahcesehir University, Istanbul, Turkey}
        \affil[2]{\normalsize Department of Electrical and Computer Engineering,
University of Maryland, MD}

        \maketitle

\begin{abstract}
We consider the age of information in G/G/1/1 systems under two service discipline models. In the first model, if a new update arrives when the service is busy, it is blocked; in the second model, a new update preempts the current update in service. For the blocking model, we first derive an exact age expression, then we propose two simple to calculate upper bounds for the average age. The first upper bound assumes  the interarrival times to have log-concave distribution. The second upper bound assumes both the interarrivals and service times to have log-concave distribution. Both upper bounds are tight in the case of M/M/1/1 systems. We show that deterministic interarrivals and service times are optimum for the blocking service model. In addition, using the age expression for G/G/1/1 systems, we calculate average age expressions for special cases, i.e., M/G/1/1 and G/M/1/1 systems. Next, for the preemption in service model, we first derive an exact average age expression for G/G/1/1 systems. Then, we propose a simple to calculate upper bound for the average age. In addition, similar to blocking discipline, using the age expression for G/G/1/1 systems, we calculate average age expressions for special cases, i.e., M/G/1/1 and G/M/1/1 systems. Average age for G/M/1/1 can be written as a summation of two terms, the first of which  depends only on the first and second moments of interarrival times and the second of which depends only on the service rate. In other words, interarrival and service times are decoupled. We show that deterministic interarrivals are optimum for G/M/1/1 systems. On the other hand, we observe for non-exponential service times that the optimal distribution of interarrival times depends on the relative values of the mean interarrival time and the mean service time.
\end{abstract}
\section{Introduction}
No matter how important information might be, there is a duration of time after which information loses its freshness. Especially in today's world of immensely interactive everything, information ages fast. Hence, in recent years, researchers have begun to consider the age of information (AoI) in addition to the value of information. Age of anything can be defined as the duration between the time of birth and the current time. This definition is sufficiently broad to cover almost all communication scenarios. However, most of the AoI literature so far has considered queueing systems with well-behaved distributions. In this paper, we take this a step forward and apply the AoI viewpoint to more general queueing distributions, and hence to more general communication scenarios.

The first papers that consider the AoI in a communication setting are\cite{Kaul12a}, \cite{Kaul12b}, and \cite{Yates12}. Reference \cite{Kaul12a} assumes First Come First Served (FCFS) systems and calculates the average AoI expressions for M/M/1, M/D/1 and D/M/1 queues, reference \cite{Kaul12b} assumes Last Come First Served (LCFS) systems with and without preemption and calculates the average AoI expression for M/M/1 queues, reference \cite{Yates12} assumes multi-source FCFS systems with M/M/1 queues, and reference \cite{Yates17a} provides a more detailed analysis. Starting with these works, there has been a growing interest in AoI analysis. For example, reference \cite{Costa16} considers a packet management approach for M/M/1/1 and M/M/1/2 queues. Reference \cite{Najm17} calculates the AoI for an M/G/1/1 queue and finds the optimum arrival rate to minimize age.

While the literature on calculating age expressions for different queueing models expands, another line of research applies the AoI approach to energy harvesting problems. The goal is to find the optimum update generation policy that minimizes age, given the service time distribution. In \cite{Sun17a}, the authors show the existence of an optimal stationary deterministic update generation policy when the service time process is a stationary and ergodic Markov chain. Application of AoI to offline energy harvesting is considered in \cite{Bacinoglu15, Arafa17a, Arafa17B}, and online energy harvesting is considered in \cite{Wu18, Arafa18a, Arafa18b, Baknina18b}.

The calculation of AoI for a given queue model requires a probabilistic approach in order to calculate the expected values of several, possibly correlated, random quantities specific to that model. The more complex the system is, the harder it is to calculate the expected values, especially when the interarrival times are not exponential. To overcome this, \cite{Yates17a} proposes an approach based on stochastic hybrid systems (SHS). In this paper, we follow an alternative path to SHS in order to analyze general cases.

In this paper, our goal is to analyze AoI for general communication scenarios, i.e., general interarrival and service time distributions. Using queueing theory terminology, our model corresponds to a G/G/1/1 system. An example of such a G/G/1/1 system is the multicast problem in \cite{Zhong17a}, where a new update is generated when a percentage of the destinations has received the current update, and service time to each destination is a shifted exponential random variable. Although an exact expression for their model is derived in \cite{Zhong17a}, in general, calculating an exact age expression for non-exponential interarrival times is difficult. For example, \cite{Buyukates18} considers a two-stage multicast extension of \cite{Zhong17a}, where only an upper bound is derived for the age of the second stage nodes. In this paper, we derive exact age expressions when the distribution of service times is arbitrary but known. In addition, we also derive upper bounds to the AoI that may be easier to use for further system design and optimization. For instance, if one designs age-minimizing policies using our upper bounds, the resulting age will be an achievable age.

We consider two service disciplines. The first one is called G/G/1/1 {\em with blocking}, where a new arrival is blocked if the server is busy. This model is also used in \cite{Costa16} for an M/M/1/1 system and in \cite{Najm17} for an M/G/1/1 system. Here, we do not restrict ourselves to exponential interarrival times or exponential service times. Our first contribution is to derive an exact expression for average age in a G/G/1/1 system. This expression can be calculated using probability density functions of interarrival and service times. Next, in order to simplify the calculations, we propose two upper bounds. The first upper bound requires the interarrival times to have log-concave distribution\footnote{Log-concave distributions have important implications for age analysis. Let us consider a random variable that corresponds to the age of a device. If the random variable has a log-concave distribution, then it has an increasing probability of failure in the next instant of time, as the device ages \cite{Marshall07}. In other words, random variables with log-concave distributions ``wear out''. As it happens, many common probability distributions that appear in arrival processes in practical systems are log-concave, including exponential, Rayleigh, Erlang, gamma with shape parameter larger than one, and uniform distributions \cite{Marshall07}.}, and the second upper bound requires both the interarrival times and the service times to have log-concave distribution.  Our upper bounds are tight for M/M/1/1 systems and they are within small percentage of the exact age for general interarrival and service time distributions. Next, we show that deterministic interarrival and service times minimize our first upper bound. In addition, we observe that deterministic interarrival times minimize the average age for a given exponential service time. Our final contribution in this service model is to calculate average age expressions for M/G/1/1 and G/M/1/1 systems. Age for M/G/1/1 systems is previously derived in \cite{Najm17}; in this paper, we provide an alternative proof using our approach. On the other hand, age  for G/M/1/1 systems is a new contribution.

Our second service discipline model is called G/G/1/1 {\em with preemption in service}, where a new arrival preempts any ongoing service. This model is used in \cite{Kaul12a} and \cite{Yates17a} for an M/M/1/1 system and in \cite{Najm17} for an M/G/1/1 system. Here, in this model as well, we do not restrict ourselves to exponential interarrival times or exponential service times. Our first contribution in this service discipline is to derive an exact expression for average age in a G/G/1/1 system. Unlike the case with blocking discipline, the average age in this model does not include any calculation of infinite sums. The age expression can be calculated relatively easily using probability density functions of interarrival and service times. In addition, we propose an even simpler upper bound expression, which does not have any stochastic ordering restrictions. Our next contribution in this service model is to calculate average age expressions for M/G/1/1 and G/M/1/1 systems. Age for M/G/1/1 systems is previously derived in \cite{Najm17}; in this paper, we provide an alternative proof using our approach. On the other hand, age for G/M/1/1 systems is a new contribution. Moreover, we prove that in a G/M/1/1 system deterministic interarrivals are optimum. We observe for non-exponential service times that the optimal distribution of interarrival times depend on the relative values of the mean interarrival time and the mean service time.

\section{System Model}

\begin{figure}
	\centering
	\subfloat[]{\includegraphics[width=.65\textwidth]{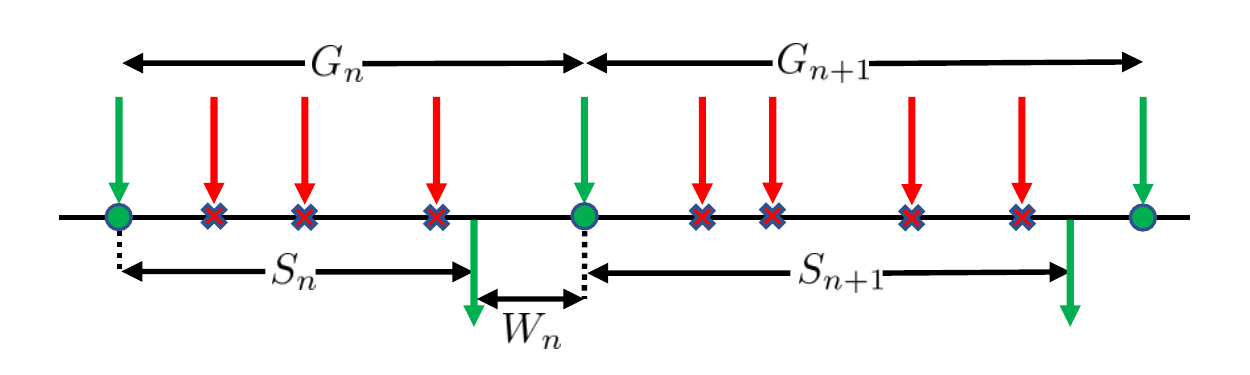}} \\
	\subfloat[]{\includegraphics[width=.65\textwidth]{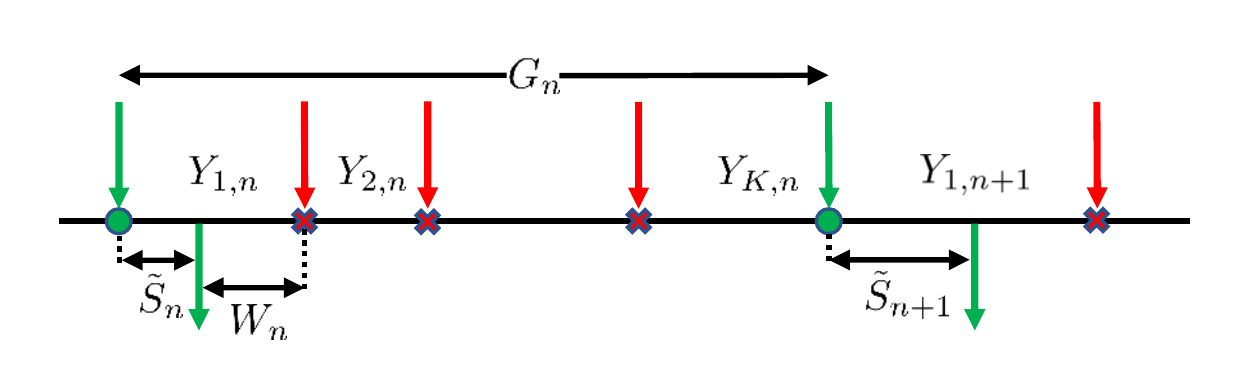}}
	\caption{Arrival and departure structure for a server. The arrows above and below the horizontal timeline corresponds to arrivals to and departures from the server. Circles are successful arrivals, while crosses are discarded arrivals: (a) blocking discipline, (b) preemption in service discipline.}
	\label{fig:geometry}
\end{figure}

We consider a communication scenario where the data arrive at the source according to an arrival process with independent and identically distributed (i.i.d.) interarrival times $Y_n$. The source transmits the data through a single server. Time duration of service is modeled as a random process with i.i.d. service times $S_n$. Interarrival times, $Y_n$, and service times, $S_n$, are independent. We specify general probability distributions for the interarrival times and service times. Fig.~\ref{fig:geometry}(a) and Fig.~\ref{fig:geometry}(b) show realizations of the arrival/departure processes with blocking and preemption in service disciplines, respectively.

\subsection{Blocking Discipline}
\label{sec:model-blocking}

In this model, if an update arrives while the server is busy, it is blocked (see cross marked arrows in Fig.~\ref{fig:geometry}(a)). If an update arrives while the server is idle, service is initiated immediately (see circle marked arrows in Fig.~\ref{fig:geometry}(a)). We refer to those updates that initiate a service as the successful updates. After a service, $S_n$, is completed, a successful update departs the server (see the arrow below the timeline in Fig.~\ref{fig:geometry}(a)). Service idle time, $W_n$, is the time between a departure of a successful update and the arrival of the next update. Interarrival times between consecutive successful updates, ${G}_n = S_n + W_n$, are called effective interarrival times. It is important to note that, the effective interarrival time, $G_n$, can be written as a random sum of random numbers, $G_n = \sum_{k=1}^K Y_{k,n}$. Note that $K$ is an integer random variable that describes the total number of arrivals before the next successful arrival. Probability mass function of $K$ can be written as
\begin{align}
\text{Pr}(K=k) = \text{Pr}\left( \sum_{j=1}^{k-1} Y_{j,n} \leq S_n < \sum_{j=1}^k Y_{j,n} \right).
\end{align}
\subsection{Preemption in Service Discipline}
\label{sec:model-preemption}

In this model, if an update arrives while the server is idle, the service is initiated immediately. If an update arrives while the server is busy, the packet being served is terminated and the new packet is pushed to the server. In Fig.~\ref{fig:geometry}(b), each arrival (arrows above the timeline) starts a service. An arrival that can finish service is called a successful arrival (see circle marked arrows in Fig.~\ref{fig:geometry}(b)). Since the service time of a successful arrival needs to be smaller than the interarrival time, the time that a successful arrival stays in service, $\tilde{S}_n$, is less than the service time of the server, i.e., $\tilde{S}_n = S_n | S_n < Y_{1,n}$.

Similar to the blocking model, interarrival times between successful updates are called effective interarrival times, $G_n$, which can be written as a random sum of random numbers, $G_n = \sum_{k=1}^K Y_{k,n}$. Unlike the blocking model, here the waiting time depends only on the current interarrival time, $W_n = Y_{1,n} - \tilde{S}_n$. Although $K$ in the blocking discipline does not follow a specific distribution, $K$ in the preemption in service discipline is a geometric random variable. An effective interarrival time is the sum of a successful arrival with probability $p=\text{Pr}(Y_{1,n}>S_n)$, and $K-1$ unsuccessful arrivals, all with the same probability $1-p$.

\section{G/G/1/1 with Blocking}

\begin{figure}[t]
     \centering
          \includegraphics[width=.6\textwidth]{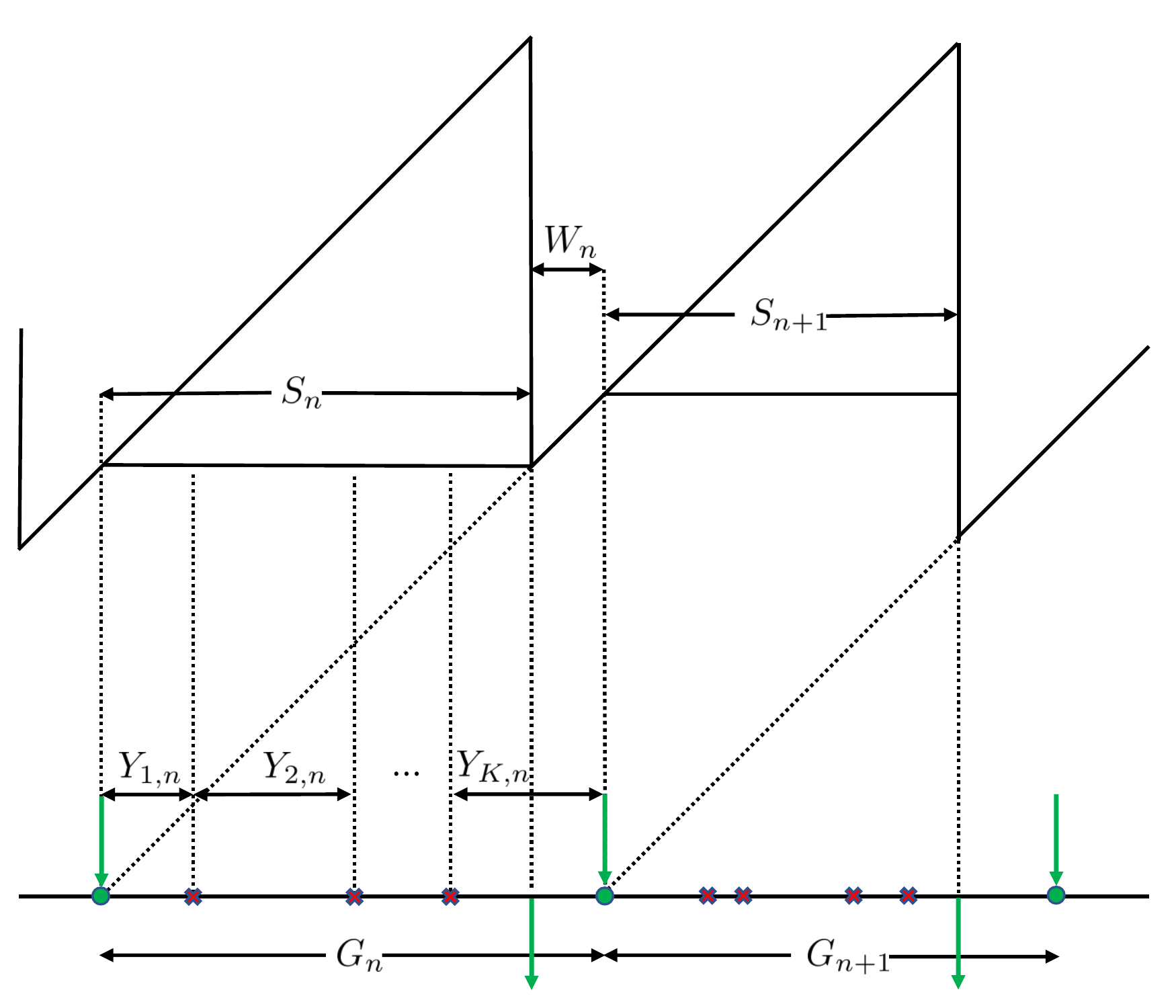}
     \caption{Age curves for G/G/1/1 with blocking model.}
     \label{fig:geometry-blocking}
\end{figure}

For G/G/1/1 with blocking discipline, average age can be written as the difference of the areas of two triangles, divided by the expected value of the effective interarrival time. From Fig.~\ref{fig:geometry-blocking}, we have
\begin{align}
\Delta_\text{G/G}^b =& \frac{E[({G}_n+S_{n+1})^2] - E[(S_{n+1})^2]}{2E[{G}_n]}   \\
=& \frac{E[{G}^2]}{2E[{G}]} + E[S],
\label{eqn:age-blocking-gg}
\end{align}
where $S_{n+1}$ is independent of $G_n$, and time indices are dropped. Next, we make a general remark about (\ref{eqn:age-blocking-gg}). The effective interarrival time, $G$, is the time of the renewal cycle of a renewal process. We see from Fig.~\ref{fig:geometry-blocking} that each time a service starts, the effective interarrival process is renewed. In \cite[page 136]{Ross96}, {\em average age of a renewal process} is defined, and is calculated as $\frac{E[G^2]}{2E[G]}$. Therefore, the average age of an information update in (\ref{eqn:age-blocking-gg}) is equal to the sum of the average age of the effective interarrival process and the average service time.

For most general interarrival and service time models, it is not easy to calculate the first and second moments of effective interarrival times $G$ needed in (\ref{eqn:age-blocking-gg}). In the section, we first derive an exact expression for (\ref{eqn:age-blocking-gg}) that depends only on the general distributions of interarrival times, $Y$, and service times, $S$. Although it is exact, the average age expression for the case of general interarrival and service time distributions require further calculations on the distribution functions of the interarrival and service times. Next, we derive simpler average age expressions for several special cases, i.e., for general interarrival and exponential service times, and exponential interarrival and general service times. Next, we go back to the case of general interarrival and service time distributions, and derive upper bounds for (\ref{eqn:age-blocking-gg}) that are easier to calculate. We observe that the tightness of these bounds depends on the parameters of the interarrival and service time distributions.

\subsection{Age for General Interarrival and Service Times}

In this section, we derive an exact age expression for the case of general interarrival and service time distributions under blocking discipline. An important aspect of this result is that the average age of an information update can be written in terms of the average age of the update arrival process, $\frac{E[Y^2]}{2E[Y]}$, instead of the average age of the effective interarrival process, $\frac{E[G^2]}{2E[G]}$.

\begin{theorem}
Consider a G/G/1/1 system with blocking discipline, where $Y_n$ are i.i.d. interarrival times with a general distribution and $S_n$ are i.i.d. service times with a general distribution. The average age of an information update in this system is
\begin{align}
\Delta_\text{G/G}^b =& \frac{E[Y^2]}{2E[Y]} + \frac{\sum_{k=1}^\infty E[A_k\bar F_S(A_k)]}{1 + \sum_{k=1}^\infty E[\bar F_S(A_k)]} + E[S]
\label{eqn:thm-blocking-gg}
\end{align}
where $A_{k} = \sum_{j=1}^{k}Y_j$, and $\bar F_S(\cdot)$ is the complementary cdf of $S$.
\label{thm:blocking-gg}
\end{theorem}

The proof of Theorem~\ref{thm:blocking-gg} is given in Section~\ref{sec:thm-blocking-gg}. Note that Theorem~\ref{thm:blocking-gg} gives the average age of a G/G/1/1 system as a summation of three terms. The first and the third terms  depend only on the interarrival and service time distributions, respectively, and the second term depends on both distributions.

\begin{figure}
	\centering
	\subfloat[]{\includegraphics[width=.33\textwidth]{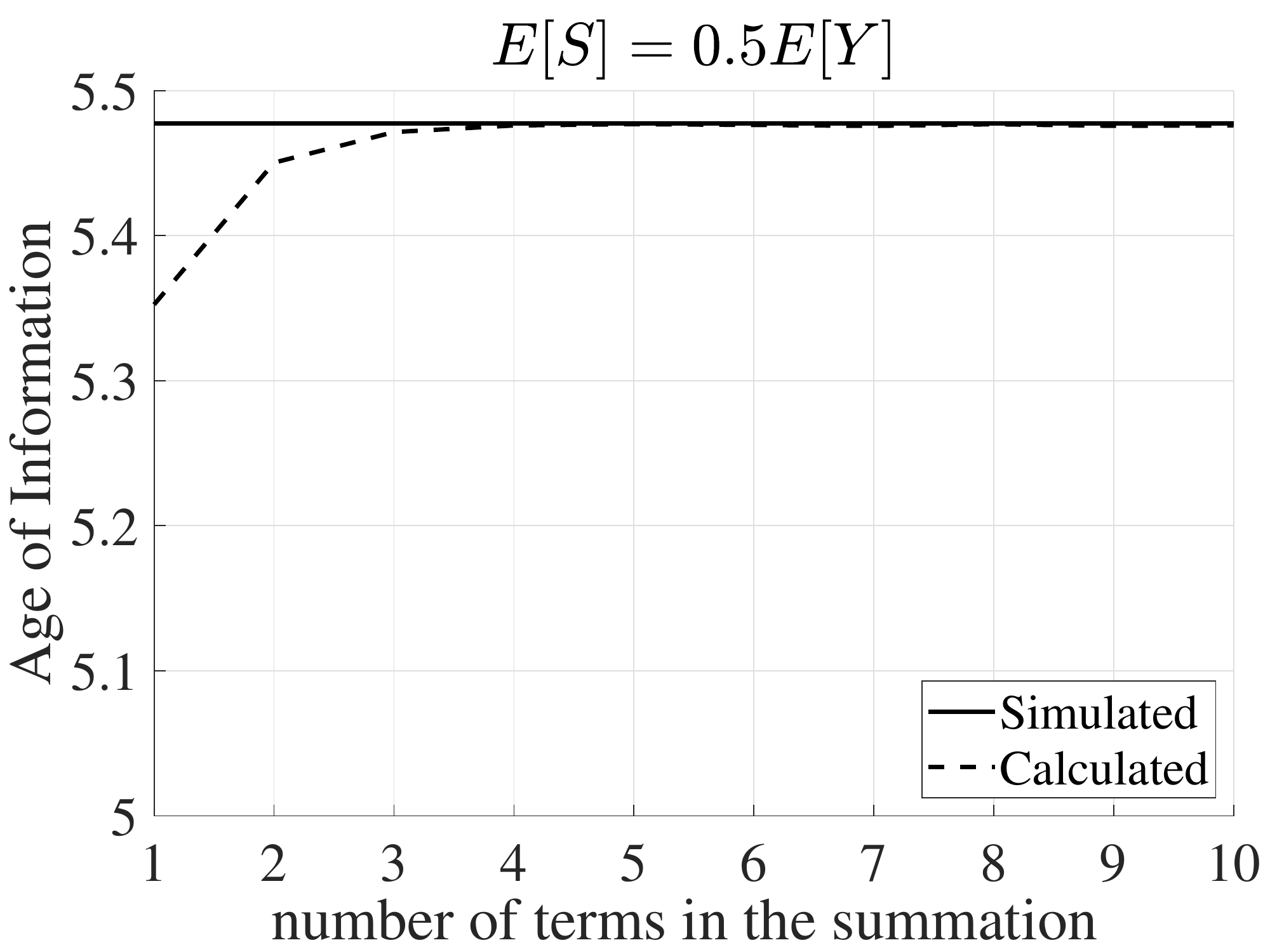}}
	\subfloat[]{\includegraphics[width=.33\textwidth]{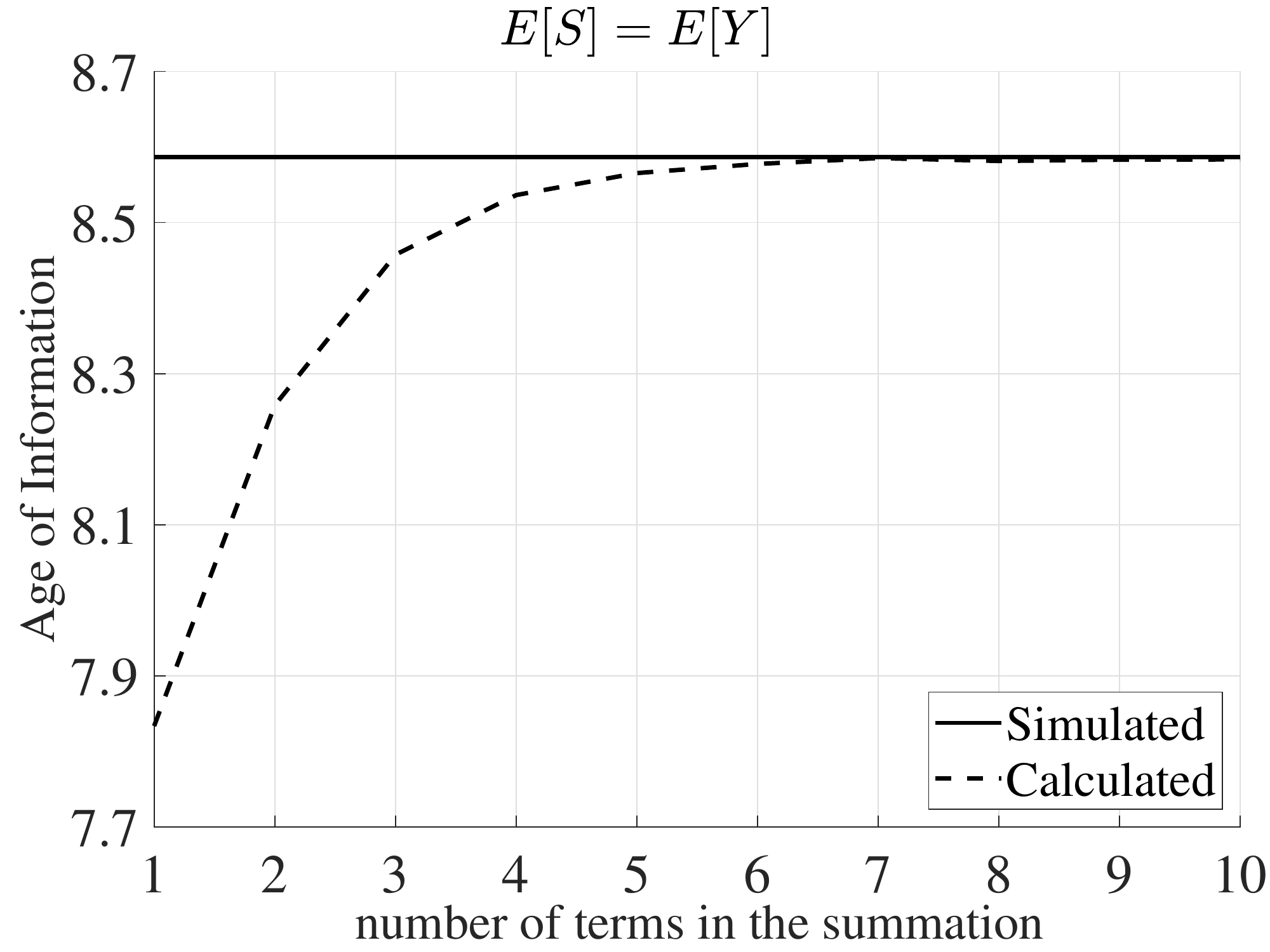}}	
	\subfloat[]{\includegraphics[width=.33\textwidth]{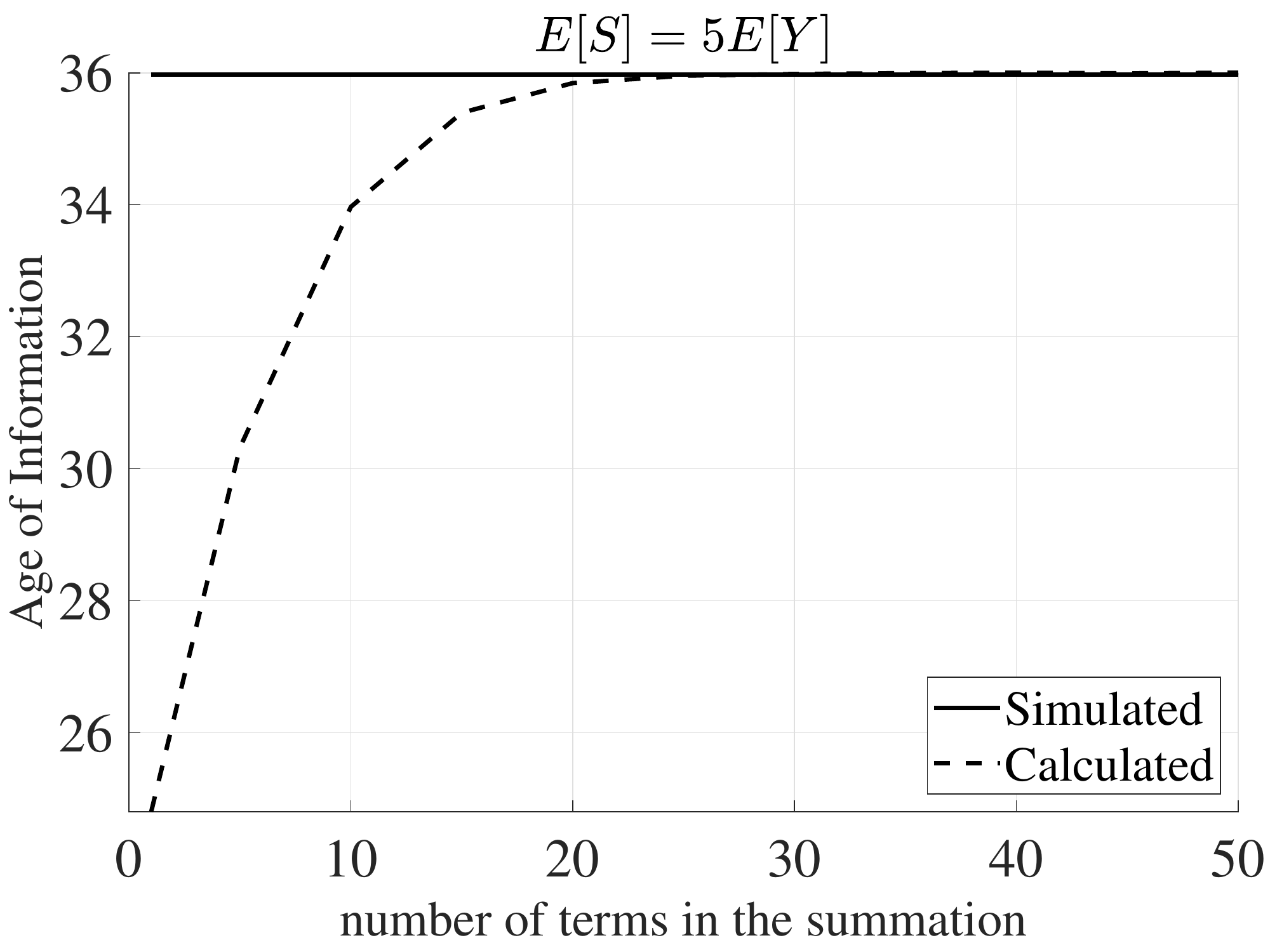}}
	\caption{Average AoI for G/G/1/1 with blocking discipline,  where the x-axis is the number of summations that are calculated in (\ref{eqn:thm-blocking-gg}) for different expected service times: (a) $E[S] = 0.5E[Y]$, (b) $E[S] = E[Y]$, (c) $E[S] = 5E[Y]$.}
	\label{fig:thm-blocking-gg}
\end{figure}

\begin{figure}
	\centering
	\subfloat[]{\includegraphics[width=.5\textwidth]{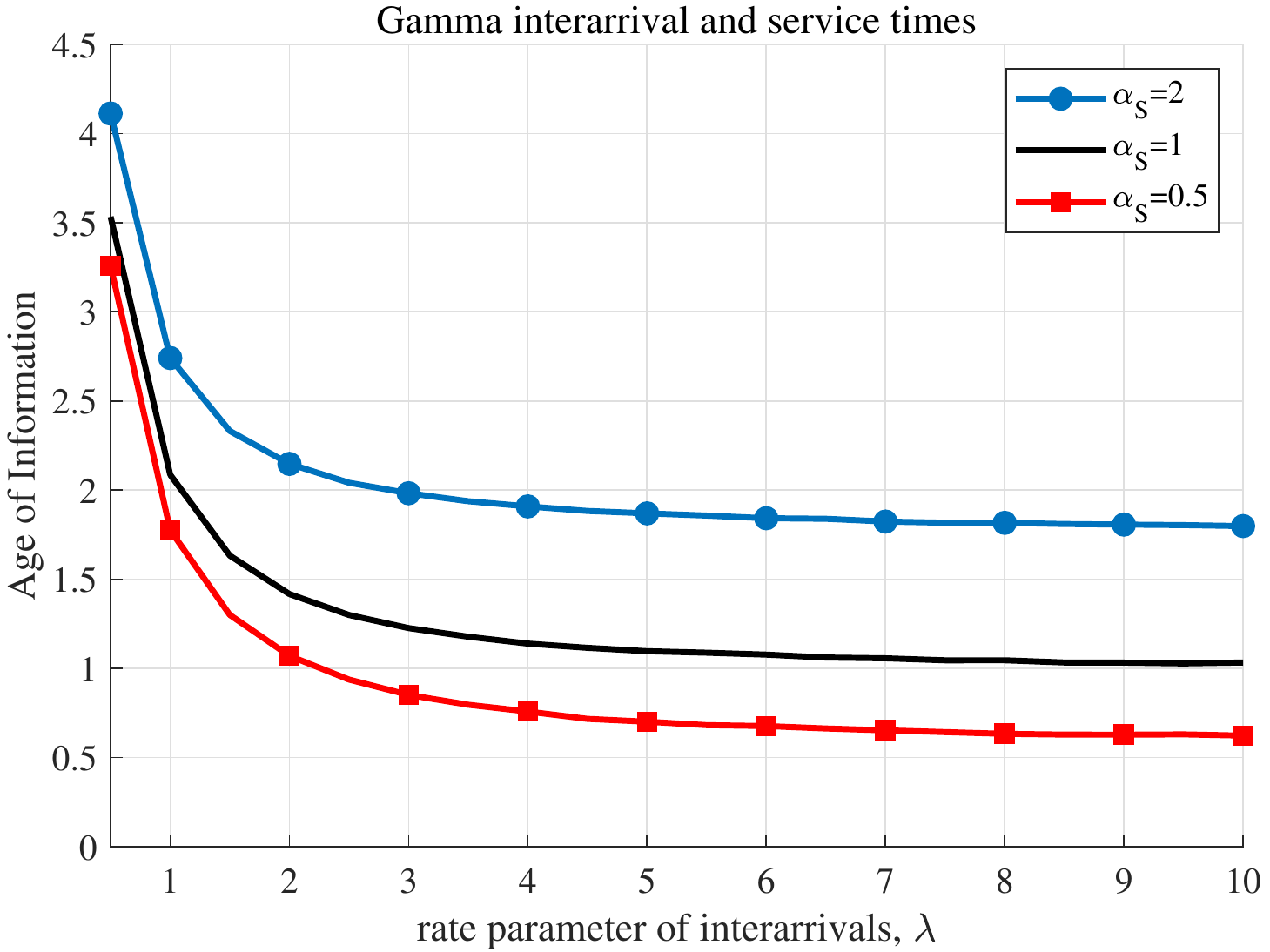}}
	\subfloat[]{\includegraphics[width=.5\textwidth]{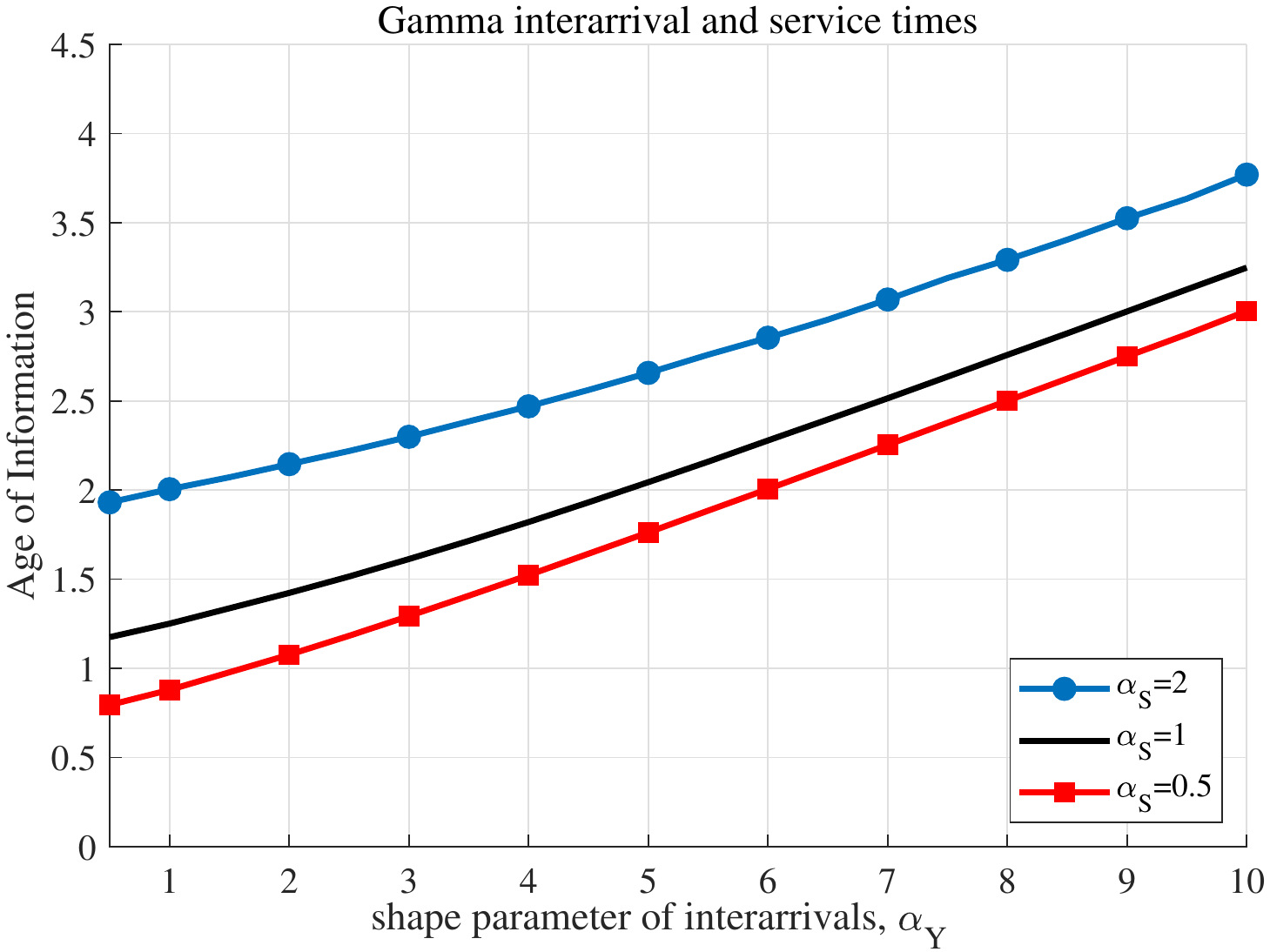}}
	\caption{Average AoI for G/G/1/1 with blocking discipline, where both the interarrival and service times are gamma distributed, (a) $\alpha_Y = \mu = 2$, (b) $\lambda = \mu = 2$.}
	\label{fig:thm-blocking-gg-rate}
\end{figure}

The exact average age expression in (\ref{eqn:thm-blocking-gg}) requires calculations of infinite sums. In Fig.~\ref{fig:thm-blocking-gg}, we consider gamma distributed\footnote{We choose gamma distribution in this paper in order to simulate general distributions. The gamma distribution forms a two-parameter exponential family. When the shape parameter of a gamma distribution is larger than one, it is log-concave; when the shape parameter is smaller than one, it is log-convex. The gamma distribution includes the chi‐squared, Erlang, and exponential distributions as special cases. Probability density function of a gamma distribution has a flexible shape so that it can be used to approximate many probabilistic models.} interarrival and service times and observe that the number of terms needed in the summation for calculation to converge depends on the ratio of the expected service time, $E[S]$, to the expected interarrival time, $E[Y]$. When $E[S] = E[Y]$, we observe from Fig.~\ref{fig:thm-blocking-gg}(b) that the error in calculation is less than 0.5\% after 5 terms in the summation. In Fig.~\ref{fig:thm-blocking-gg-rate}, we consider gamma distributed interarrival times where $\alpha_Y$ is the shape parameter and $\lambda$ is the rate parameter, and gamma distributed service times where  $\alpha_S$ is the shape parameter and $\mu$ is the rate parameter. We observe the effects of the rate and shape parameters of gamma distribution on the average age. We observe that the age decreases monotonically with the rate parameter and  increases monotonically with the shape parameter of either distribution.

\subsection{Age for General Interarrival and Exponential Service Times}

For exponential service times, we have a closed form expression for the average age. First, we show that for exponentially distributed service times with rate parameter $\mu$, $K$ is a geometric random variable with $p = 1-E[e^{-\mu Y}]$.
\begin{lemma}
Consider a G/M/1/1 system with blocking discipline, where $Y_n$ are i.i.d. interarrival times with a general distribution, $S_n$ are i.i.d. service times with an exponential distribution with rate parameter $\mu$, and $K_n$ are i.i.d.  discrete random variables with
\begin{align}
{\rm Pr}(K=k) = {\rm Pr}\left( \sum_{j=1}^{k-1} Y_j \leq S < \sum_{j=1}^k Y_j \right).
\label{eqn:lem-blocking-gm-geometric}
\end{align}
Then, K is geometric with $p = 1-E[e^{-\mu Y}]$.
\label{lem:blocking-gm-geometric}
\end{lemma}

The proof of Lemma~\ref{lem:blocking-gm-geometric} is given in Section~\ref{sec:lem-blocking-gm-geometric}. Using this lemma, in the next theorem, we derive the average age of a G/M/1/1 system for a given distribution for the interarrival times.

\begin{theorem}
\label{thm:blocking-gm}
Consider a G/M/1/1 system with blocking discipline, where $Y_n$ are i.i.d. interarrival times with a general distribution and $S_n$ are i.i.d. service times with an exponential distribution with rate parameter $\mu$. The average age of this system is
\begin{align}
\Delta_\text{G/M}^b =& \frac{E[Y^2]}{2E[Y]} + \frac{E[Ye^{-\mu Y}]}{1-E[e^{-\mu Y}]} + \frac{1}{\mu}
\label{eqn:thm-blocking-gm}
\end{align}
\end{theorem}

The proof of Theorem~\ref{thm:blocking-gm} is given in Section~\ref{sec:thm-blocking-gm}. Note that similar to Theorem~\ref{thm:blocking-gg} in G/G/1/1 case, Theorem~\ref{thm:blocking-gm} gives the average age of a G/M/1/1 system as a summation of three terms. The first and the third terms  depend only on the interarrival and service time distributions, respectively, and the second term depends on both distributions (note the second term has $\mu$ in it even though the expectation is with respect to $Y$). Unlike the G/G/1/1 case, the middle term in the average age of a G/M/1/1 system in (\ref{eqn:thm-blocking-gm}) does not contain any summations. Therefore, the average age of a G/M/1/1 system can be calculated rather easily given the distribution of interarrival times. For example, when the interarrival times are also exponential with rate parameter $\lambda$, (\ref{eqn:thm-blocking-gm}) reduces to
\begin{align}
\Delta_\text{M/M}^p = & \frac{1}{\lambda} + \frac{2}{\mu} - \frac{1}{\lambda+\mu}
\end{align}
which is the age of an M/M/1/1 queue that is found in \cite{Costa16}.

In the next corollary, we propose an equivalent average age expression for a G/M/1/1 system with blocking discipline given in Theorem~\ref{thm:blocking-gm}.
\begin{corollary}
\label{cor:thm-blocking-gm-equiv}
The average age in Theorem~\ref{thm:blocking-gm} can be written equivalently as
\begin{align}
\Delta_\text{G/M}^b =& \frac{E[Y^2]}{2E[Y]} + 2E[S] - E[S | S < Y].
\label{eqn:cor-thm-blocking-gm-equiv}
\end{align}
\end{corollary}

The proof of Corollary~\ref{cor:thm-blocking-gm-equiv} is given in Section~\ref{sec:cor-thm-blocking-gm-equiv}. When $Y$ is exponential, this corollary directly gives the average age of an M/M/1/1 system, since the random variable $S | S < Y$ is exponential with rate parameter $\lambda+\mu$ when $Y$ and $S$ are exponential with parameters $\lambda$ and $\mu$, respectively. However, we observe that calculating the average age for a general $Y$ using Theorem~\ref{thm:blocking-gm} is easier.

\begin{figure}
	\centering
	\subfloat[]{\includegraphics[width=.5\textwidth]{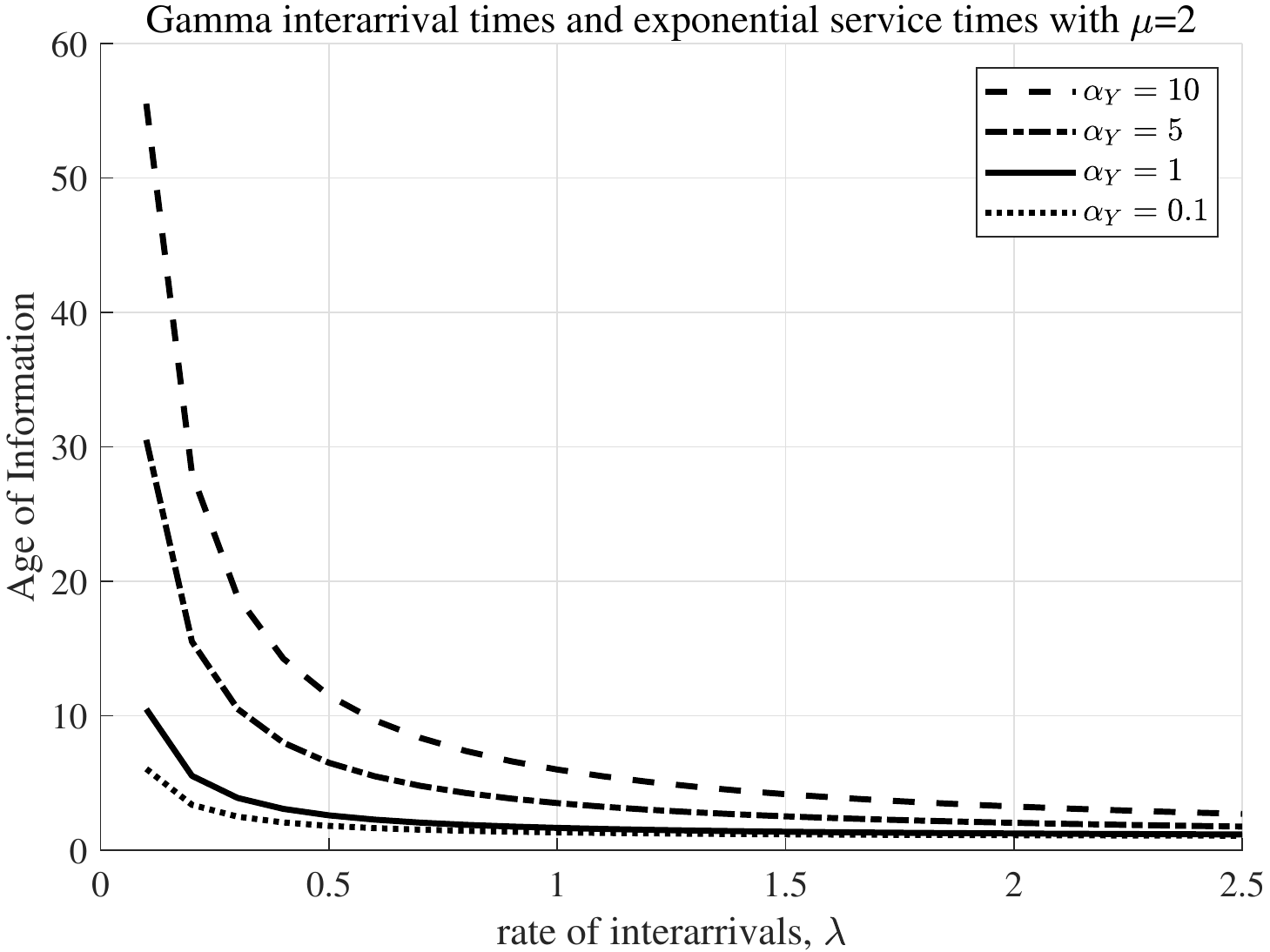}}
	\subfloat[]{\includegraphics[width=.5\textwidth]{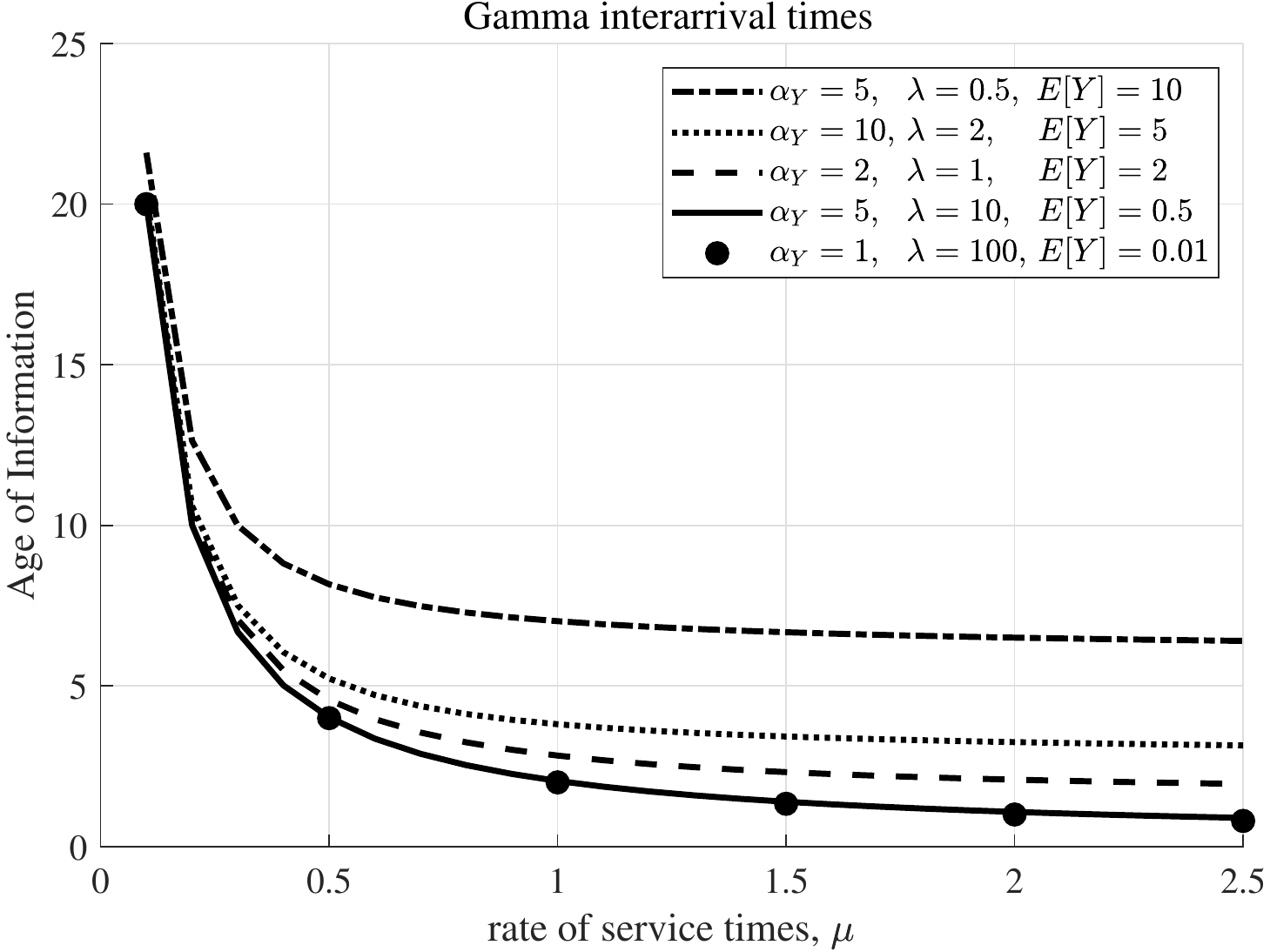}}
	\caption{Average AoI for G/M/1/1 with blocking discipline (a) with respect to $\lambda$ when $\mu=2$ and for several $\alpha$, (b) with respect to $\mu$ for several ($\alpha$, $\lambda$) pairs.}
	\label{fig:thm-blocking-gm}
\end{figure}

In Fig.~\ref{fig:thm-blocking-gm}, we consider gamma distributed interarrival times where $\alpha_Y$ is the shape parameter and $\lambda$ is the rate parameter, and exponential service times where $\mu$ is the rate parameter. In Fig.~\ref{fig:thm-blocking-gm}(a), we plot the average age with respect to $\lambda$ when $\alpha_Y$ and $\mu$ are fixed, and in Fig.~\ref{fig:thm-blocking-gm}(b), we plot the average age with respect to $\mu$ when $\alpha_Y$ and $\lambda$ are fixed. We observe that the age decreases with both rate parameters. Fig.~\ref{fig:thm-blocking-gm}(a) shows that smaller $\alpha_Y$ results in a lower age. We see from Fig.~\ref{fig:thm-blocking-gm}(b) that the lowest age is achieved when the mean interarrival time is the smallest.

\begin{figure}
	\centering
	\subfloat{\includegraphics[width=.5\textwidth]{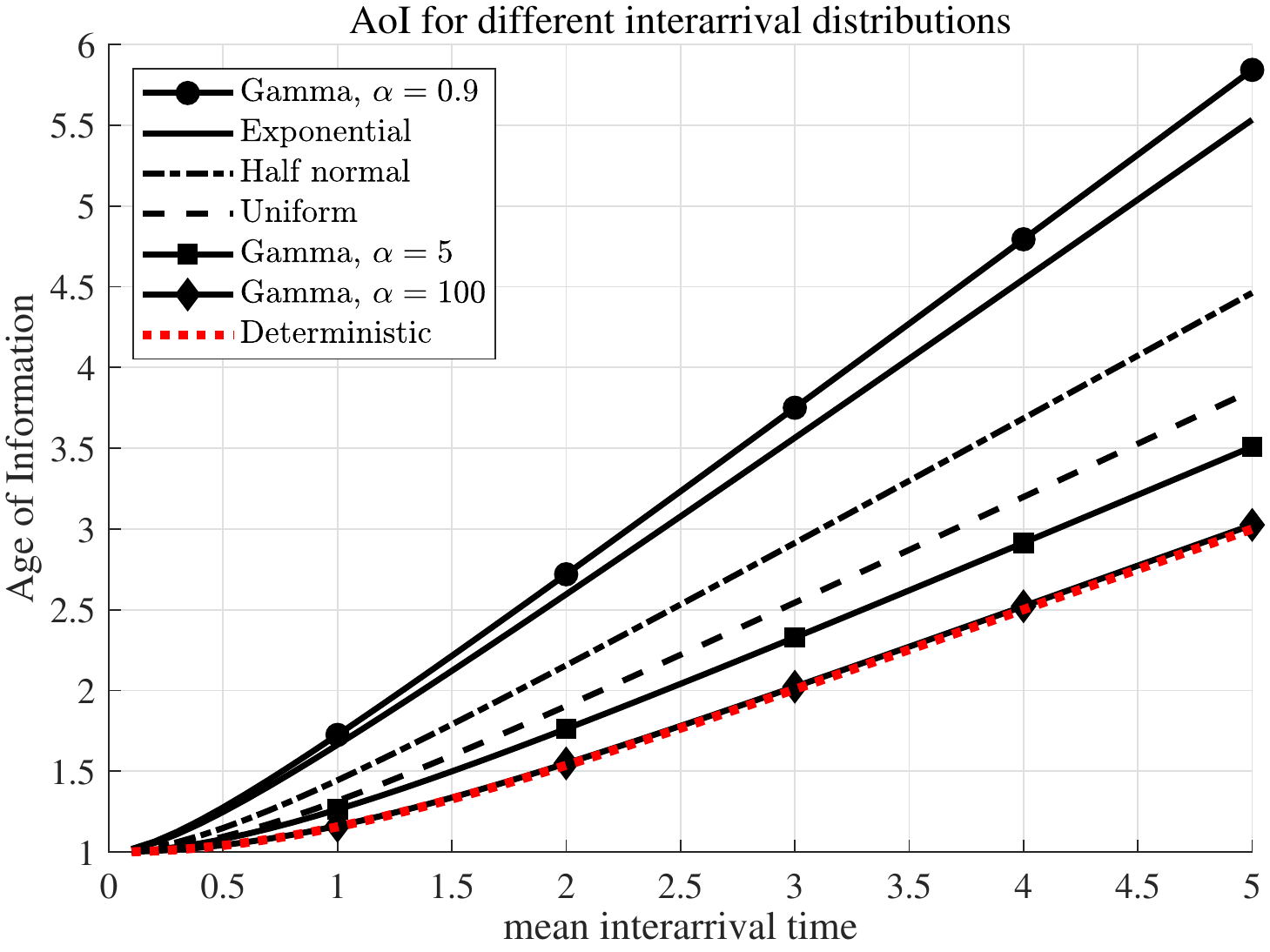}}
	\caption{Average AoI for G/M/1/1 with blocking discipline with respect to mean interarrival time for several interarrival distributions.}
	\label{fig:thm-blocking-gm-2}
\end{figure}

At this point, it is natural to ask what the age minimizing interarrival time distribution is for a G/M/1/1 system with a given mean interarrival time. The first term on the right hand side of (\ref{eqn:thm-blocking-gm}) suggests that a distribution with the smallest second moment, which belongs to a deterministic random variable, would minimize the age. However, the effect of the middle term  on the right hand side of (\ref{eqn:thm-blocking-gm})  is not immediately clear. In Fig.~\ref{fig:thm-blocking-gm-2}, we plot  (\ref{eqn:thm-blocking-gm}) for several distributions. We observe that deterministic interarrivals result in the minimum age for a given mean expected interarrival time. In addition, we observe that exponential interarrivals are the worst among log-concave distributions in terms of the resulting age. Note that gamma distributions with $\alpha < 1$ are not log-concave.

\subsection{Age for Exponential Interarrival and General Service Times}

In this section, we consider exponential interarrival and general service times, i.e., an M/G/1/1 system, with blocking discipline. This system is considered in \cite{Najm17} as well. Let us give the outline of the derivation in \cite{Najm17}  starting from (\ref{eqn:age-blocking-gg}). Note that when interarrivals are exponential, the waiting time after a service is completed is also exponential. Therefore $G =Y + S$. Then, using the independence of $Y$ and $S$, we have
\begin{align}
\Delta_\text{M/G}^b =& \frac{E[(Y+S)^2]}{2(E[Y]+E[S])} + E[S] \\
=& \frac{\frac{2}{\lambda^2}+\frac{4E[S]}{\lambda} + 2E^2[S] +E[S^2] }{2(\frac{1}{\lambda}+E[S])}\\
=& \frac{1}{\lambda} + \frac{ \lambda E[S^2] }{2(1+ \lambda E[S])} +E[S]
\label{eqn:blocking-mg}
\end{align}
where $E[Y] = \frac{1}{\lambda}$ and $E[Y^2] = \frac{2}{\lambda^2}$. In Theorem~\ref{thm:blocking-mg} below, we provide an alternative proof to (\ref{eqn:blocking-mg}) that calculates the age expression in Theorem~\ref{thm:blocking-gg} for an exponential $Y$. Although this is a replication of a previous result, Theorem~\ref{thm:blocking-mg} shows how our general G/G/1/1 result in Theorem~\ref{thm:blocking-gg} can be used for specific cases. In addition, the proof method here will be useful in proving Theorem~\ref{thm:blocking-lcg}.

\begin{theorem}
\label{thm:blocking-mg}
Consider an M/G/1/1 system with blocking discipline, where $Y_n$ are i.i.d. exponential interarrival times with rate parameter $\lambda$ and $S_n$ are i.i.d. service times with a general distribution. The average age of this system is
\begin{align}
\Delta_\text{M/G}^b =& \frac{1}{\lambda} + \frac{\lambda E[S^2]}{2\left(1+\lambda E[S]\right)} + E[S]
\label{eqn:thm-blocking-mg}
\end{align}
\end{theorem}

The proof of Theorem~\ref{thm:blocking-mg} is given in Section~\ref{sec:thm-blocking-mg}. We refer the reader to \cite{Najm17} for further analysis of M/G/1/1 systems. In the next section, we generalize the result for M/G/1/1 systems to LC/G/1/1 systems where the interarrival times have a log-concave distribution.

\subsection{Age for Log-Concave Interarrival and General Service Times}

Our results in the previous sections provide exact age expressions for single server and single packet in the system queues with blocking discipline under different assumptions on the distributions of interarrival and service times. Theorem~\ref{thm:blocking-gg} gives the most general result for a G/G/1/1 system. Although exact, the age expression in (\ref{eqn:thm-blocking-gg}) requires calculation of an infinite sum. In this section, we derive an upper bound to (\ref{eqn:thm-blocking-gg}), which is easy to calculate when the distribution of interarrival times is log-concave. As discussed before, most distributions we encounter in queueing systems are log-concave.

\begin{theorem}
\label{thm:blocking-lcg}
Consider an LC/G/1/1 system with blocking discipline, where $Y_n$ are i.i.d. interarrival times with a log-concave distribution, and $S_n$ are i.i.d. service times with a general distribution. An upper bound for the average age of this system is
\begin{align}
\Delta_\text{LC/G}^b \leq & \frac{E[Y^2]}{2E[Y]} + \frac{E[S^2]}{2E[K]E[Y]} + E[S]
\label{eqn:thm-blocking-lcg}
\end{align}
\end{theorem}

The proof of Theorem~\ref{thm:blocking-lcg} is given in Section~\ref{sec:thm-blocking-lcg}. This theorem provides an upper bound for the average age of a large class of interarrival and service time distributions. Theorem~\ref{thm:blocking-lcg} is useful for service time distributions where calculating $E[K]$ is rather easy. For example, for an exponential service time, the exact age expression in Theorem~\ref{thm:blocking-gm} can be upper bounded by an expression given in Corollary~\ref{cor:blocking-lcm} below.

\begin{corollary}
\label{cor:blocking-lcm}
Consider an LC/M/1/1 system with blocking discipline, where $Y_n$ are i.i.d. interarrival times with a log-concave distribution, and $S_n$ are i.i.d. exponential service times with rate parameter $\mu$. An upper bound to the average age of this system is
\begin{align}
\Delta_\text{LC/M}^d \leq \frac{E[Y^2]}{2E[Y]} + \frac{1-E[e^{-\mu Y}]}{\mu^2 E[Y]} + \frac{1}{\mu}
\label{eqn:cor-blocking-lcm}
\end{align}
\end{corollary}

The proof of Corollary~\ref{cor:blocking-lcm} follows from (\ref{eqn:thm-blocking-lcg}) by noting that $E[S] = \frac{1}{\mu}$ and $E[S^2] = \frac{2}{\mu^2}$ for an exponential $S$, and $E[K] = \frac{1}{1 - E[e^{-\mu Y}]}$ for a geometric $K$. In order to evaluate the upper bound in (\ref{eqn:cor-blocking-lcm}), one only needs to know the first and second moments of $Y$ and its moment generating function. On the other hand, for service time distributions where calculating $E[K]$ is not easy, we have another upper bound given in Corollary~\ref{cor:blocking-lcg} below, which further upper bounds Theorem~\ref{thm:blocking-lcg}.

\begin{corollary}
\label{cor:blocking-lcg}
Consider an LC/G/1/1 system with blocking discipline, where $Y_n$ are i.i.d. interarrival times with a log-concave distribution, and $S_n$ are i.i.d. service times with a general distribution. An upper bound to the average age of this system is
\begin{align}
\Delta_\text{LC/G}^b \leq & \frac{E[Y^2]}{2E[Y]} + \frac{E[S^2]}{2E[S]} + E[S]
\label{eqn:cor-blocking-lcg}
\end{align}
\end{corollary}

The proof of Corollary~\ref{cor:blocking-lcg} follows by remembering from Section~\ref{sec:model-blocking} that $G = W + S$. Since $W$ is non-zero, we have $E[G] = E[K]E[Y] > E[S]$. It is interesting to note that, in the upper bound of Corollary~\ref{cor:blocking-lcg}, the first term  depends only on the first and second moments of the interarrival times and the second and the third terms  depend only on the first and second moments of the service times. In other words, interarrival and service times are decoupled. The upper bound to the AoI is equal to the sum of the age of the interarrival process, $\frac{E[Y^2]}{2E[Y]}$, the age of the service process, $\frac{E[S^2]}{2E[S]}$, and the mean service time, $E[S]$.

\begin{figure}
	\centering
	\subfloat[]{\includegraphics[width=.5\textwidth]{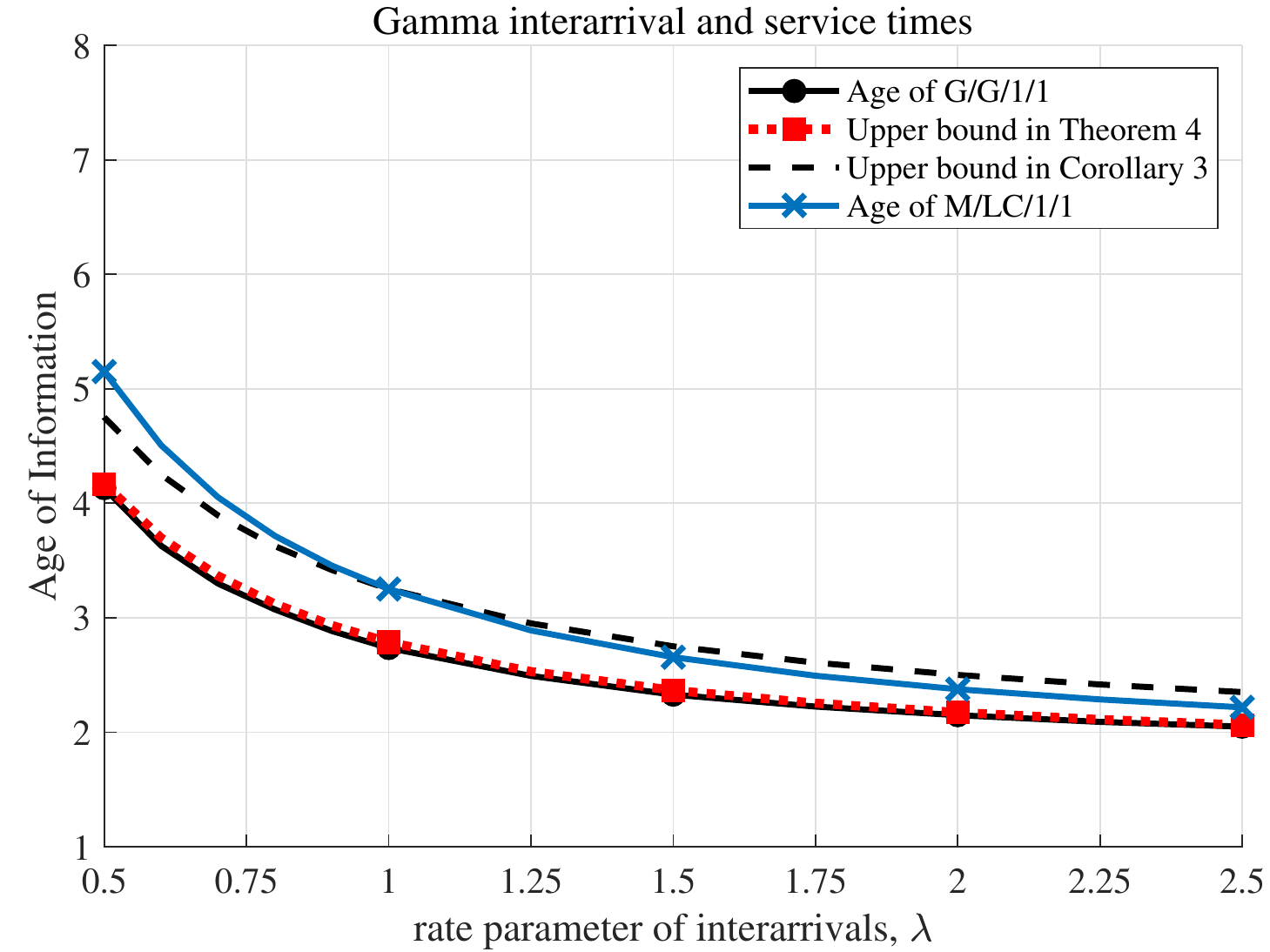}}
	\subfloat[]{\includegraphics[width=.5\textwidth]{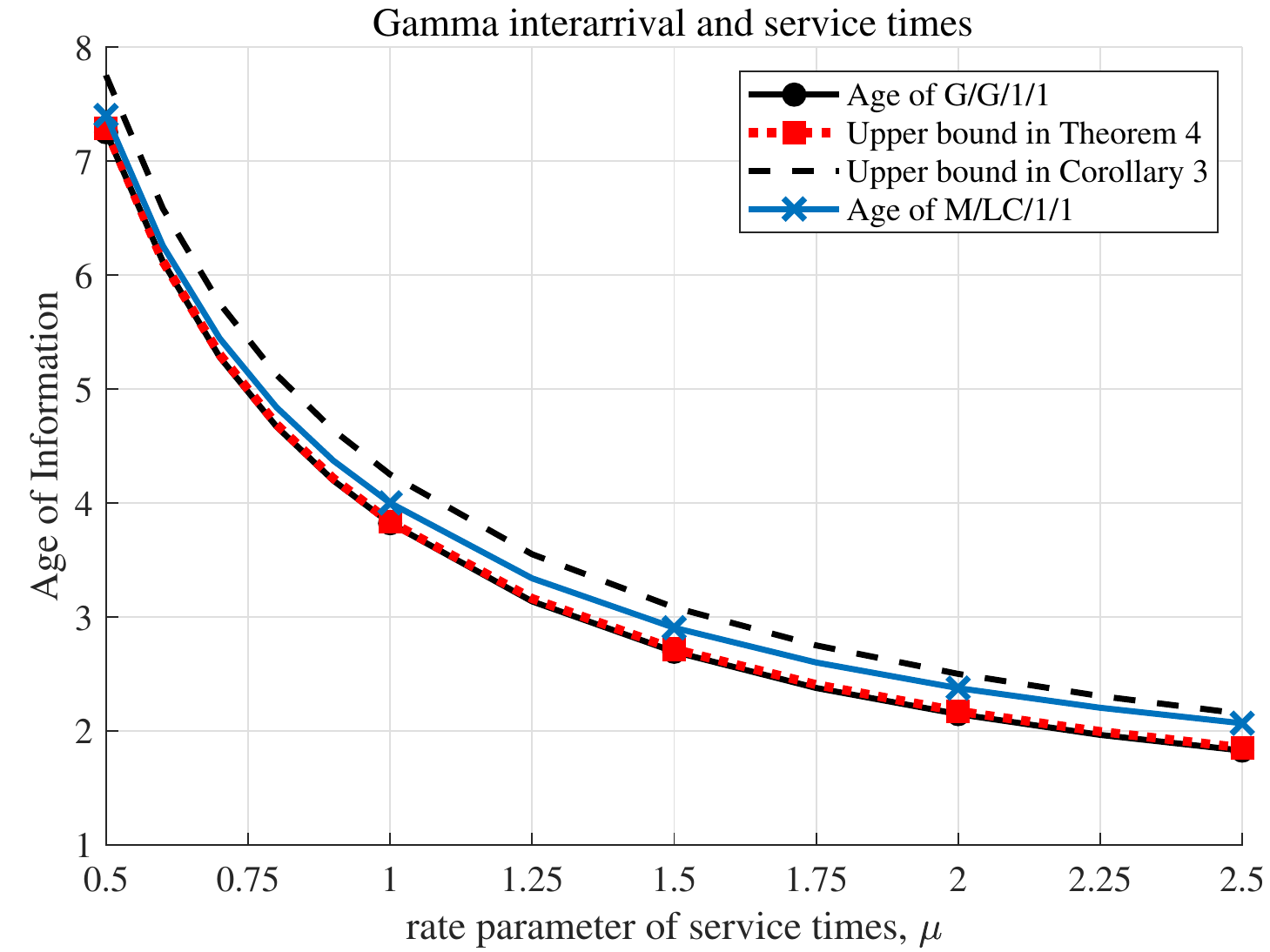}}
	\caption{Average AoI for an LC/G/1/1 system  with blocking discipline, where interarrival and service times are gamma distributed; (a) AoI with respect to rate parameter of interarrivals, $\lambda$, when $\mu = \alpha_Y = \alpha_S = 2$, (a) AoI with respect to rate parameter of service times, $\mu$, when $\lambda = \alpha_Y = \alpha_S = 2$.}
	\label{fig:thm-blocking-ng-1}
\end{figure}

\begin{figure}
	\centering
	\subfloat[]{\includegraphics[width=.5\textwidth]{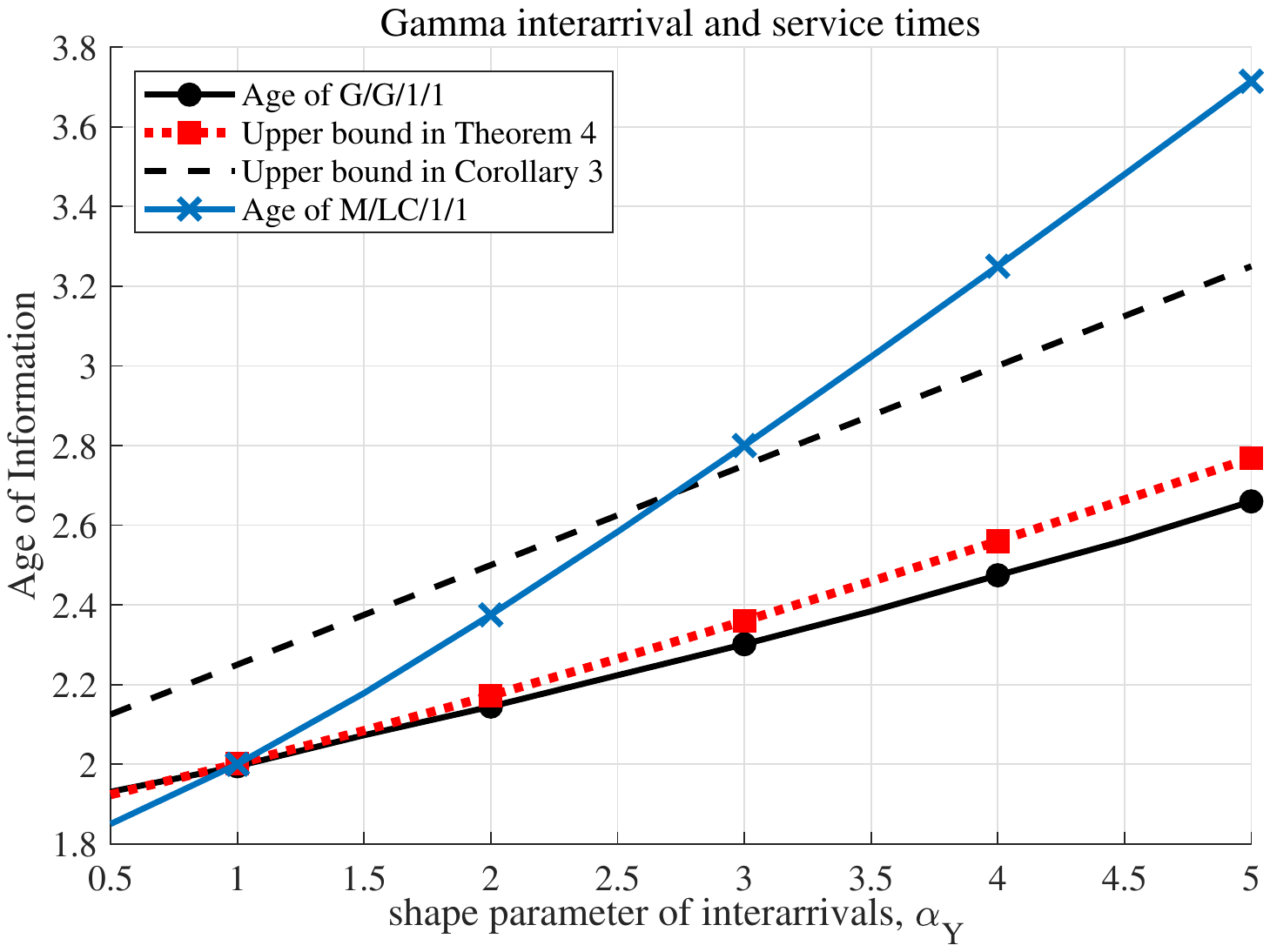}}
	\subfloat[]{\includegraphics[width=.5\textwidth]{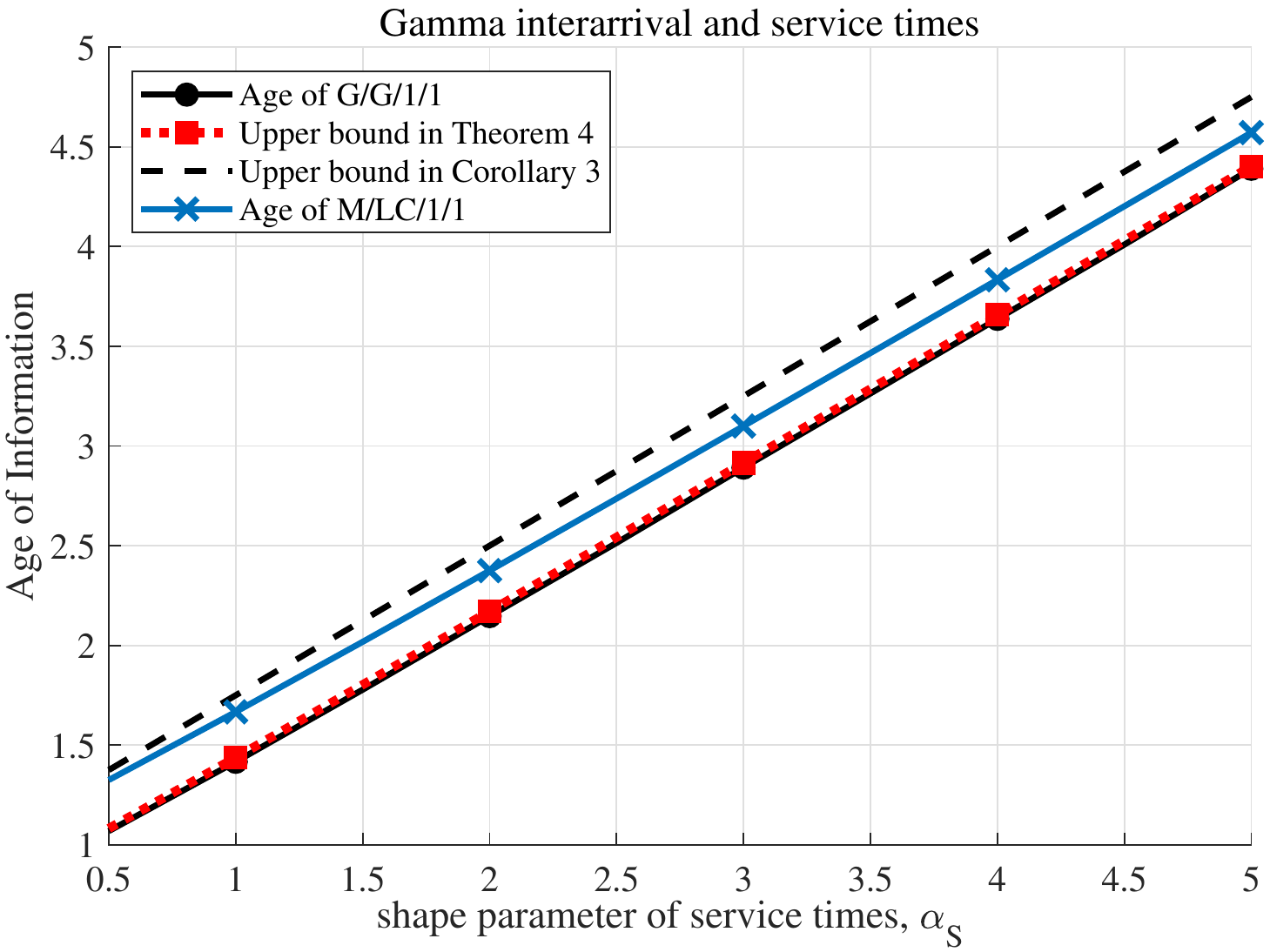}}
		\caption{Average AoI for an LC/G/1/1 system  with blocking discipline, where interarrival and service times are gamma distributed; (a) AoI with respect to shape parameter of interarrivals, $\alpha_Y$, when $\mu = \lambda = \alpha_S = 2$, (a) AoI with respect to shape parameter of service times, $\alpha_S$, when $\lambda = \alpha_Y = \mu = 2$.}
	\label{fig:thm-blocking-ng-2}
\end{figure}

In Figs.~\ref{fig:thm-blocking-ng-1} and \ref{fig:thm-blocking-ng-2}, we examine the tightness of our bounds, where solid, dotted and dashed lines correspond to the simulated exact age, the upper bound in Theorem~\ref{thm:blocking-lcg}, and the upper bound in  Corollary~\ref{cor:blocking-lcg}, respectively. Lines with crosses will be introduced later in the text. Both the interarrival times and the service times follow a gamma distribution which is log-concave when its shape parameter is larger than one. We observe that the upper bound given in Theorem~\ref{thm:blocking-lcg} is almost the same as the exact age while the upper bound given in Corollary~\ref{cor:blocking-lcg} has an almost constant difference from the exact age for a large set of parameter values. In Fig.~\ref{fig:thm-blocking-ng-2}(a), the expression on the right hand side of (\ref{eqn:thm-blocking-lcg}) is not an upper bound for $\alpha_Y <1$, since gamma distribution is no longer log-concave for $\alpha_Y <1$. On the other hand, (\ref{eqn:thm-blocking-lcg}) holds for $\alpha_S < 1$ as seen from Fig.~\ref{fig:thm-blocking-ng-2}(b).

Since the upper bound in  (\ref{eqn:cor-blocking-lcg}) follows the exact age closely, in Corollary~\ref{cor:blocking-lcg-opt} below we examine the probability distributions that minimize and maximize this upper bound under given fixed mean interarrival and service time constraints.

\begin{corollary}
\label{cor:blocking-lcg-opt}
Consider an LC/G/1/1 system with blocking discipline, where $Y_n$ are i.i.d. interarrival times with a log-concave distribution, and $S_n$ are i.i.d. service times with a general distribution. Given mean interarrival and service time constraints, deterministic intearrival and service times minimize the upper bound in (\ref{eqn:cor-blocking-lcg}), and exponential interarrival times maximize  the upper bound in (\ref{eqn:cor-blocking-lcg}).
\end{corollary}
\begin{Proof}
For a given $E[Y]$ and $E[S]$, (\ref{eqn:cor-blocking-lcg}) is minimized when the second moments, or equivalently, the variances of $Y$ and $S$ are minimized. Since deterministic variables have zero variance, deterministic intearrival and service times minimize the upper bound in (\ref{eqn:cor-blocking-lcg}).

When $Y$ has a log-concave distribution, we have the following relation between the first and second moments of $Y$ from \cite[Proposition 6.A.6]{Marshall07}
\begin{align}
E[Y^2] \leq 2 E^2[Y]
\label{eqn:prf-cor-blocking-lcg-op-01}
\end{align}
where the equality is achieved with an exponential distribution. Therefore, we conclude that exponential interarrivals result in the largest possible $\frac{E[Y^2]}{2E[Y]}$.
\end{Proof}

Although the objective function of the minimization in Corollary~\ref{cor:blocking-lcg-opt} is an upper bound for the average age, we recall that a similar conclusion is drawn for the exact age in Fig.~\ref{fig:thm-blocking-gm-2}.

Corollary~\ref{cor:blocking-lcg-opt} suggests that the upper bound for the age becomes the largest if the interarrival time distribution is replaced by an exponential distribution with the same mean. Note that this is not the same as the age of an M/G/1/1 system. The right hand side of (\ref{eqn:thm-blocking-mg}) is different (larger) than the right hand side of (\ref{eqn:cor-blocking-lcg}) with exponential $Y$.  In the next section, we show under certain conditions that the age of G/G/1/1 is upper bounded by the age of M/G/1/1 with the same mean interarrival time.

\subsection{Age for Log-Concave Interarrival and Log-Concave Service Times}

In this section, we derive another upper bound for the age of G/G/1/1 systems, which requires the interarrival times and the service times to have log-concave distribution. Similar to (\ref{eqn:cor-blocking-lcg}), this upper bound  depends only on the first and second moments of the interarrival and service time distributions.

\begin{theorem}
Consider a LC/LC/1/1 system with blocking discipline, where $Y_n$ are i.i.d. interarrival times with a log-concave distribution, and $S_n$ are i.i.d. service times with a log-concave distribution. Also consider an M/LC/1/1 system that is formed by replacing the interarrival times of LC/LC/1/1 system with exponentially distributed interarrival times, $Y^e_n$, where $E[Y^e]=E[Y]$. Then, the average age of the LC/LC/1/1 system, $\Delta_\text{LC/LC}^b$ is upper bounded by the average age of M/LC/1/1 system, $\Delta_\text{M/LC}^b$.
\label{thm:blocking-dn}
\end{theorem}

The proof of Theorem~\ref{thm:blocking-dn} is given in Section~\ref{sec:thm-blocking-dn}. Note that gamma distribution is log-concave when its shape parameter is $\alpha > 1$. In Figs.~\ref{fig:thm-blocking-ng-1} and \ref{fig:thm-blocking-ng-2}, both the interarrival times and the service times follow a gamma distribution and $\Delta_\text{M/LC}^b$ curves are plotted using a cross mark.  We observe that when the rate parameter of interarrivals is large or the shape parameter of the interarrivals is small, the upper bound  in Theorem~\ref{thm:blocking-dn} is tighter than that in Corollary~\ref{cor:blocking-lcg}. In fact, the age of M/LC/1/1 system converges to the age of LC/LC/1/1 system with increasing $\lambda$ and decreasing $\alpha_Y$. In Fig.~\ref{fig:thm-blocking-ng-2}(a), the age in M/LC/1/1 is not an upper bound for $\alpha_Y <1$, since gamma distribution is no longer log-concave for $\alpha_Y <1$.

\section{G/G/1/1 with preemption in service}

\begin{figure}[t]
     \centering
          \includegraphics[width=.6\textwidth]{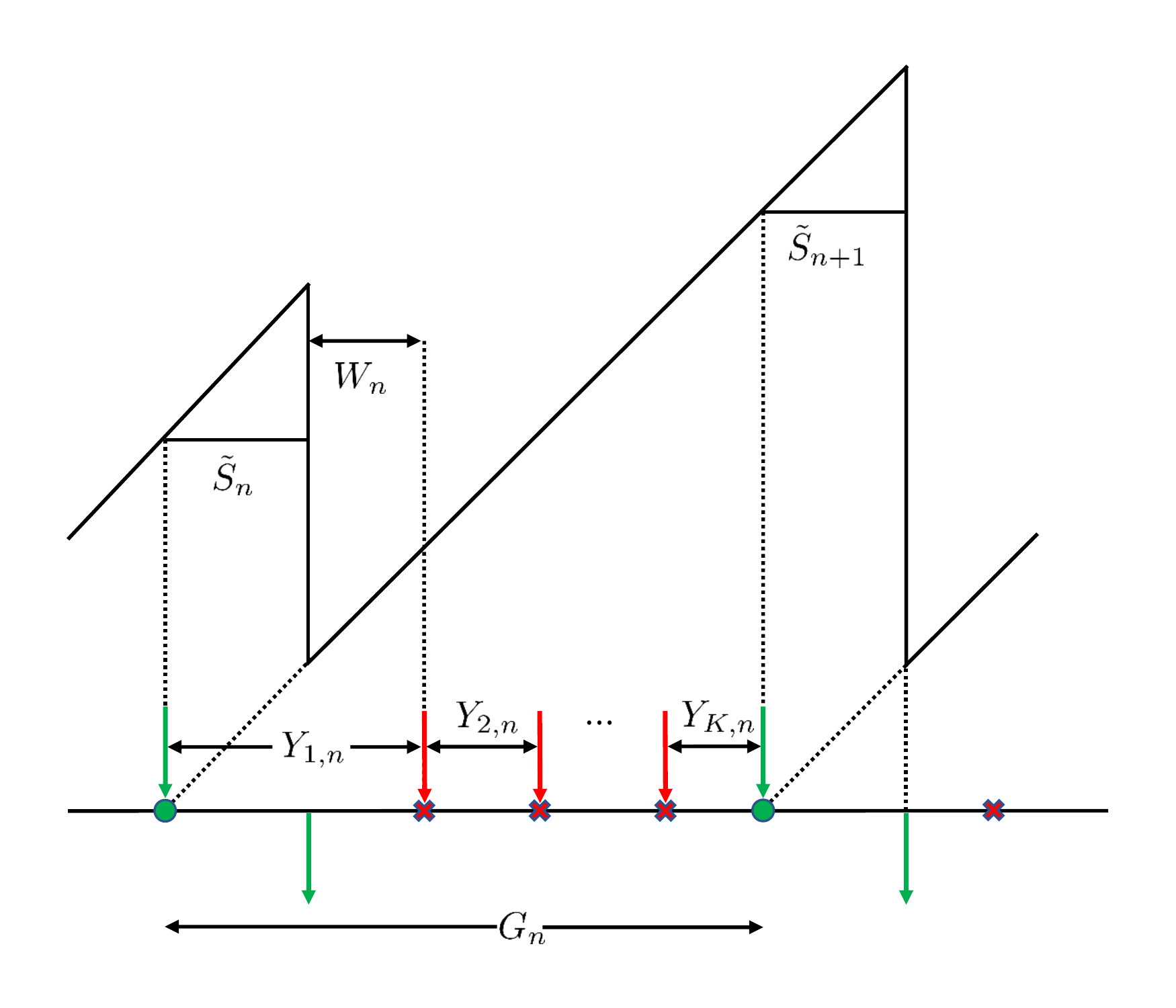}
     \caption{Age curves for G/G/1/1 with preemption in service.}
     \label{fig:geometry-preemption}
\end{figure}

Similar to the case with blocking discipline, for G/G/1/1 with preemption in service discipline as well, average age can be written as the difference of the areas of two triangles, divided by the expected length of the effective interarrival time. From Fig.~\ref{fig:geometry-preemption}, we have
\begin{align}
\Delta_\text{G/G}^p =& \frac{E[({G}_n+\tilde{S}_{n+1})^2] - E[(\tilde{S}_{n+1})^2]}{2E[{G}_n]}   \\
=& \frac{E[{G}^2]}{2E[{G}]} + E[\tilde{S}]
\label{eqn:age-preemption}
\end{align}
where $\tilde{S}_{n+1} = \{S|S<Y\}_{n+1}$ is independent of $G_n$, and time indices are dropped.

It is important to note that the random variable $G$ in this model is not the same as the $G$ in the blocking model. The difference can be observed from Figs.~\ref{fig:geometry-blocking} and \ref{fig:geometry-preemption} by noting the change in scale for $S_n$. However, $G$ represents the effective interarrival time and $\frac{E[G^2]}{2E[G]}$ represents the average age of effective interarrival process \cite[page 136]{Ross96} in this model as well. Therefore, the average age of an information update in (\ref{eqn:age-preemption}) is the sum of the average age of the effective interarrival process and the amount of time update spends in service, which is different than the mean service time of the server.

In this section, we first derive an exact closed form expression for (\ref{eqn:age-preemption}) that  depends only on the general distributions of interarrival times, $Y$, and service times, $S$. Unlike the age expression in the blocking model, the age expression in the preemption in service model does not require further calculations. Next, we derive simpler average age expressions for special cases, i.e., for general interarrival and exponential service times, and exponential interarrival and general service times. Since the exact age expressions are already simple to calculate in the preemption in service discipline, we derive only a single upper bound for the case of general interarrival and service time distributions.

\subsection{Age for General Interarrival and Service Times}

In this section, we derive an exact age expression for the case of general interarrival and service time distributions under the preemption in service discipline. An important aspect of this result is that, unlike the case with blocking discipline, the average age of an information update in this model does not involve any infinite sums. In addition, similar to the case with blocking discipline, the average age of an information update can be written in terms of the average age of the update arrival process, $\frac{E[Y^2]}{2E[Y]}$, instead of the average age of the effective interarrival process, $\frac{E[G^2]}{2E[G]}$.

\begin{theorem}
Consider a G/G/1/1 system with preemption in service discipline, where $Y_n$ are i.i.d. interarrival times with a general distribution and $S_n$ are i.i.d. service times with a general distribution. The average age of an information update in this system is
\begin{align}
\Delta_\text{G/G}^p =& \frac{E[Y^2]}{2E[Y]} + \frac{E[Y \bar F_S(Y)]}{1-E[\bar F_S(Y)]} + E[\tilde{S}]
\label{eqn:thm-preemption-gg}
\end{align}
where $\bar F_S(\cdot)$ is the complementary cdf of $S$, and $\tilde{S}\!= \!S|S<Y$.
\label{thm:preemption-gg}
\end{theorem}

The proof of Theorem~\ref{thm:preemption-gg} is given in Section~\ref{sec:thm-preemption-gg}. Similar to Theorem~\ref{thm:blocking-gg} in the case of blocking discipline, Theorem~\ref{thm:preemption-gg} in the case of preemption in service discipline gives the average age of a G/G/1/1 system as a summation of three terms. However, unlike Theorem~\ref{thm:blocking-gg} which includes infinite sums, Theorem~\ref{thm:preemption-gg} is much easier to calculate given the distributions of interarrival and service times. Yet, an even simpler expression that is given in Corollary~\ref{cor:preemption-gg-ub} below can be used to upper bound the age in Theorem~\ref{thm:preemption-gg}.

\begin{corollary}
Consider a G/G/1/1 system with preemption in service, where $Y_n$ are interarrival times and $S_n$ are service times. The average age of this system is upper bounded by
\begin{align}
\Delta_\text{G/G}^p \leq& \frac{E[Y^2]}{2E[Y]} + \frac{E[Y] E[\bar F_S(Y)]}{1-E[\bar F_S(Y)]} + E[\tilde{S}]
\label{eqn:cor-preemption-gg-ub}
\end{align}
where $\bar F_S(\cdot)$ is the complementary cdf of $S$, and $\tilde{S} \!=\! S|S<Y$.
\label{cor:preemption-gg-ub}
\end{corollary}

The proof of Corollary~\ref{cor:preemption-gg-ub} follows by noting that $E[Y|Y<S] \leq E[Y]$. The upper bound in (\ref{eqn:cor-preemption-gg-ub}) is tight when $K$ is independent of $Y_k$. An example of this is the multicast model in \cite{Zhong17a}, where the random sum parameter $K$ is independent of $Y_k$.

\begin{figure}
	\centering
	\subfloat{\includegraphics[width=.5\textwidth]{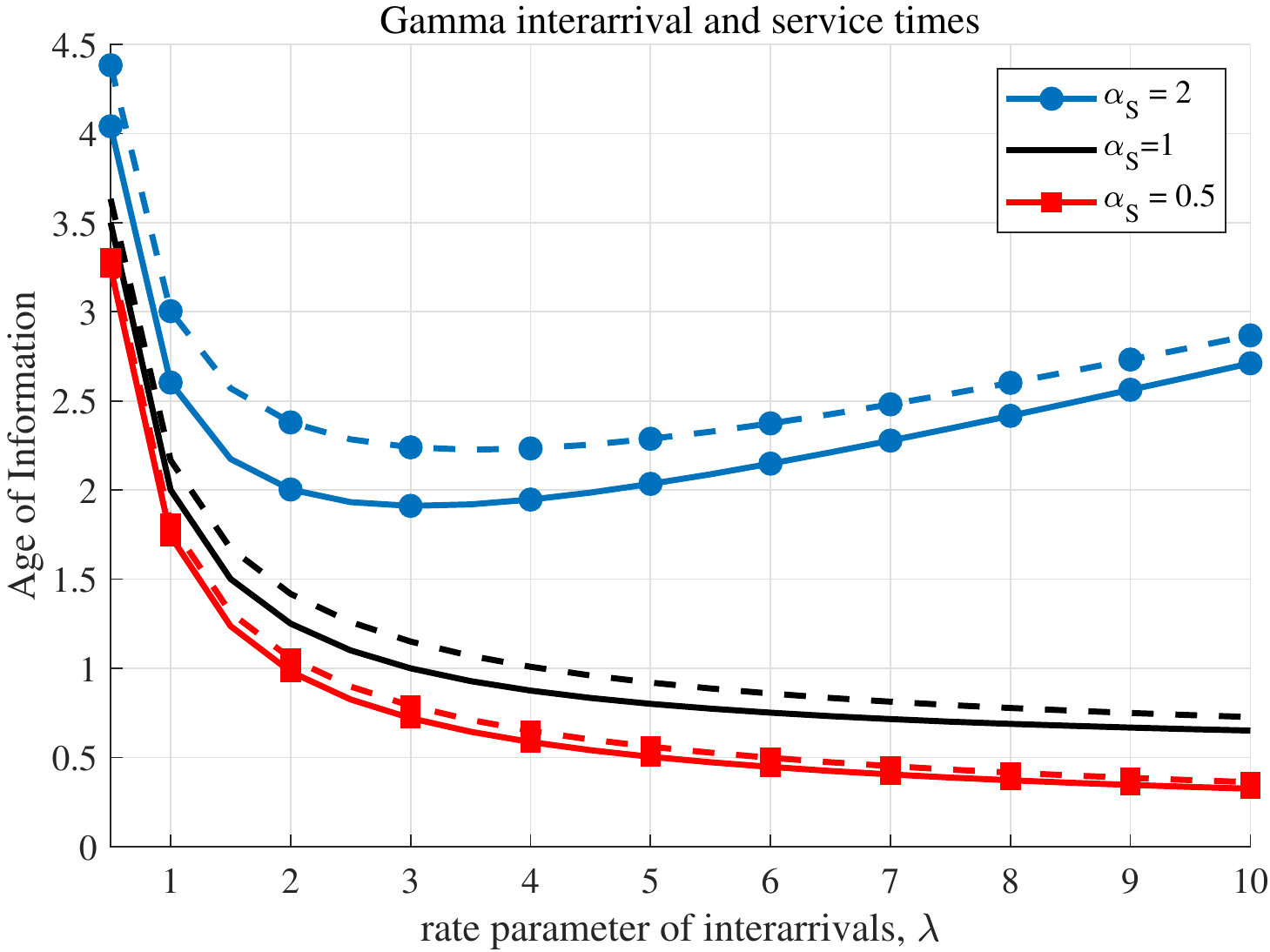}}
	\subfloat{\includegraphics[width=.5\textwidth]{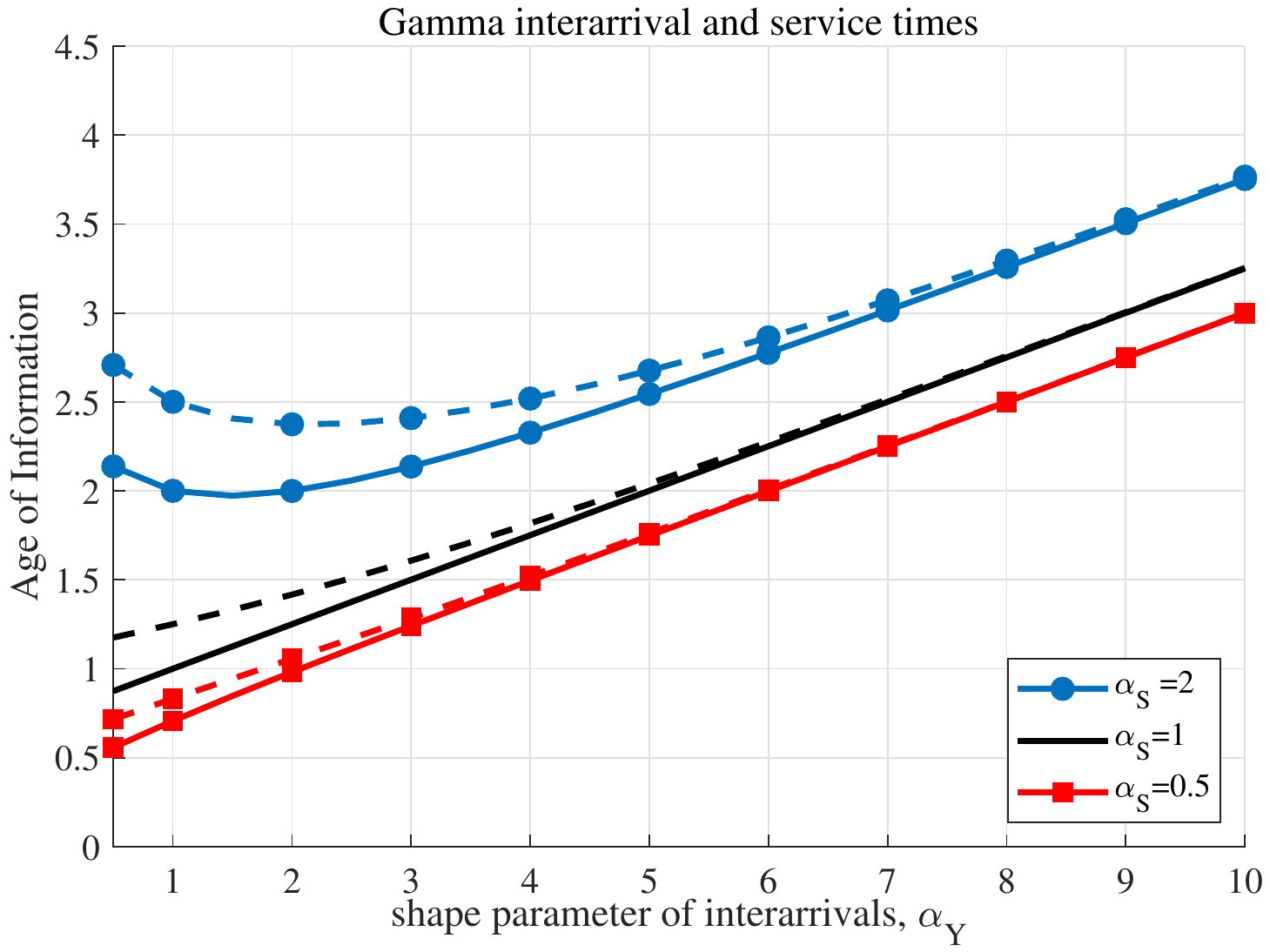}}
	\caption{Average AoI for G/G/1/1 with preemption in service discipline, where both the interarrival and service times are gamma distributed. Solid lines correspond to exact age expressions and dashed lines corresponds to upper bounds.}
	\label{fig:thm-preemption-gg}
\end{figure}

In order to examine the tightness of the bound in Corollary~\ref{cor:preemption-gg-ub} for a general case, in Fig.~\ref{fig:thm-preemption-gg}, we simulate the same G/G/1/1 system as in the case with blocking discipline, calculate its age using Theorem~\ref{thm:preemption-gg} (solid lines in Fig.~\ref{fig:thm-preemption-gg}) and compare it to the upper bound in Corollary~\ref{cor:preemption-gg-ub} (dashed lines in Fig.~\ref{fig:thm-preemption-gg}). We observe that the difference between the exact age and the upper bound is bounded and small. We also observe that the difference between the exact age and the upper bound depends on the interarrival and service time distributions. In addition, we observe that the age in the preemption in service discipline is affected by the distribution of the service time more than the age in the blocking discipline.

We have seen from Fig.~\ref{fig:thm-blocking-gg-rate} through Fig.~\ref{fig:thm-blocking-ng-2} that the age in the blocking discipline is monotone with respect to the parameters of the gamma distribution. This is not the case in the preemption in service discipline which suggests that age minimizing distribution is not deterministic for a G/G/1/1 system. When $\lambda$ is very large, in other words when the interarrivals are too frequent, preemption starts to overload the system when service time distribution is log-concave, i.e., $\alpha_S > 1$. Time duration between two successive successful interarrivals gets larger, and hence age increases. This observation for G/G/1/1 systems with preemption in service differs significantly from M/M/1/1 systems with preemption in service, where age is monotonically decreasing in $\lambda$ \cite{Yates17a} (see also the unmarked curves in Fig.~\ref{fig:thm-preemption-gg}(a)). In addition, the minimum age for G/G/1/1 systems over the rate parameter is smaller in the blocking scenario than it is in the preemption in service scenario. However, we know from \cite{Yates17a} and \cite{Costa16} that the opposite is true for M/M/1/1 systems (see also the unmarked curves in Figs.~\ref{fig:thm-blocking-gg-rate}(a) and \ref{fig:thm-preemption-gg}(a)). These observations reassure our initial motivation to consider the AoI for G/G/1/1 systems, as they can behave significantly different than M/M/1/1 systems.

\begin{figure}
	\centering
	\subfloat{\includegraphics[width=.5\textwidth]{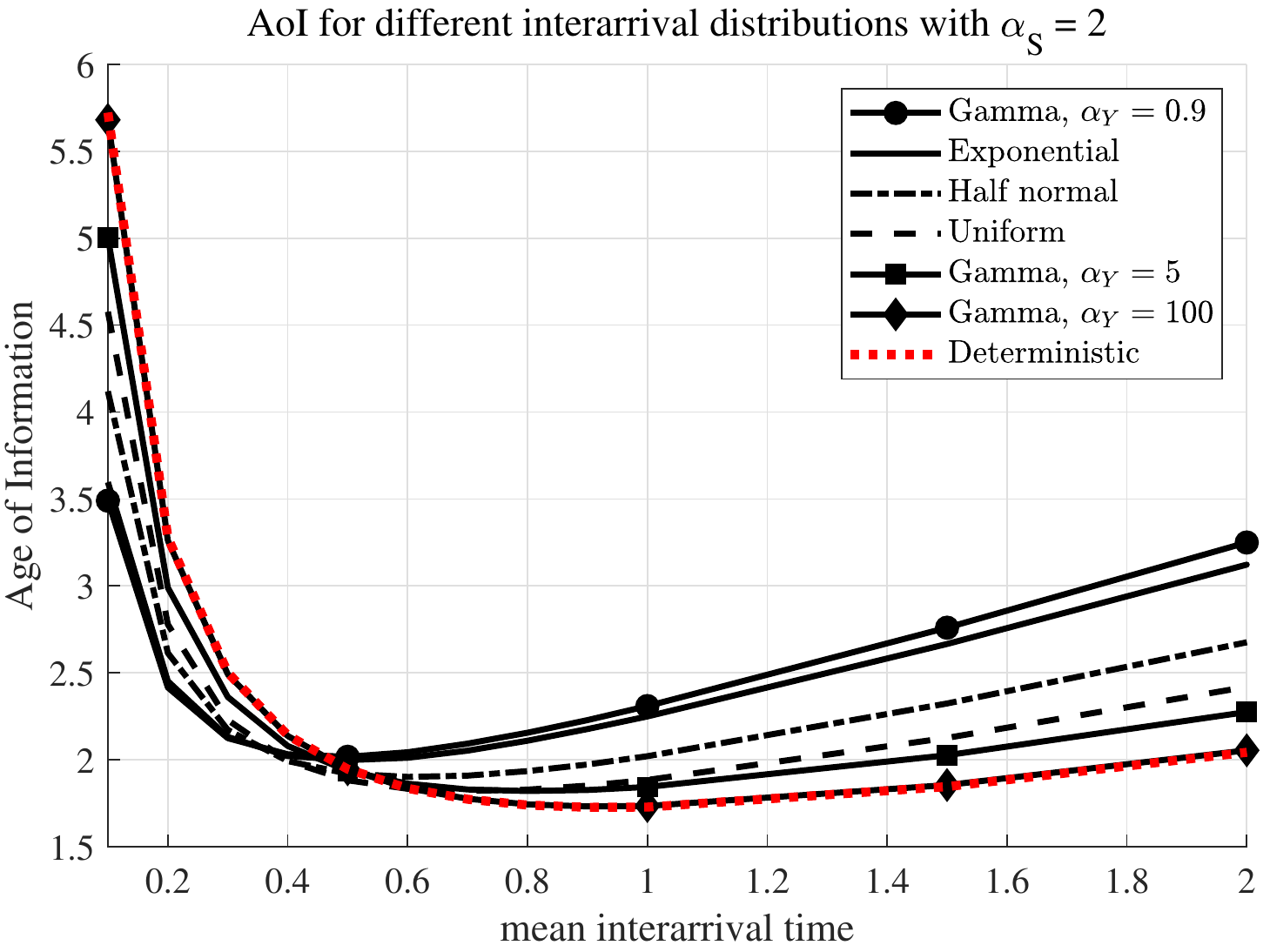}}
	\subfloat{\includegraphics[width=.5\textwidth]{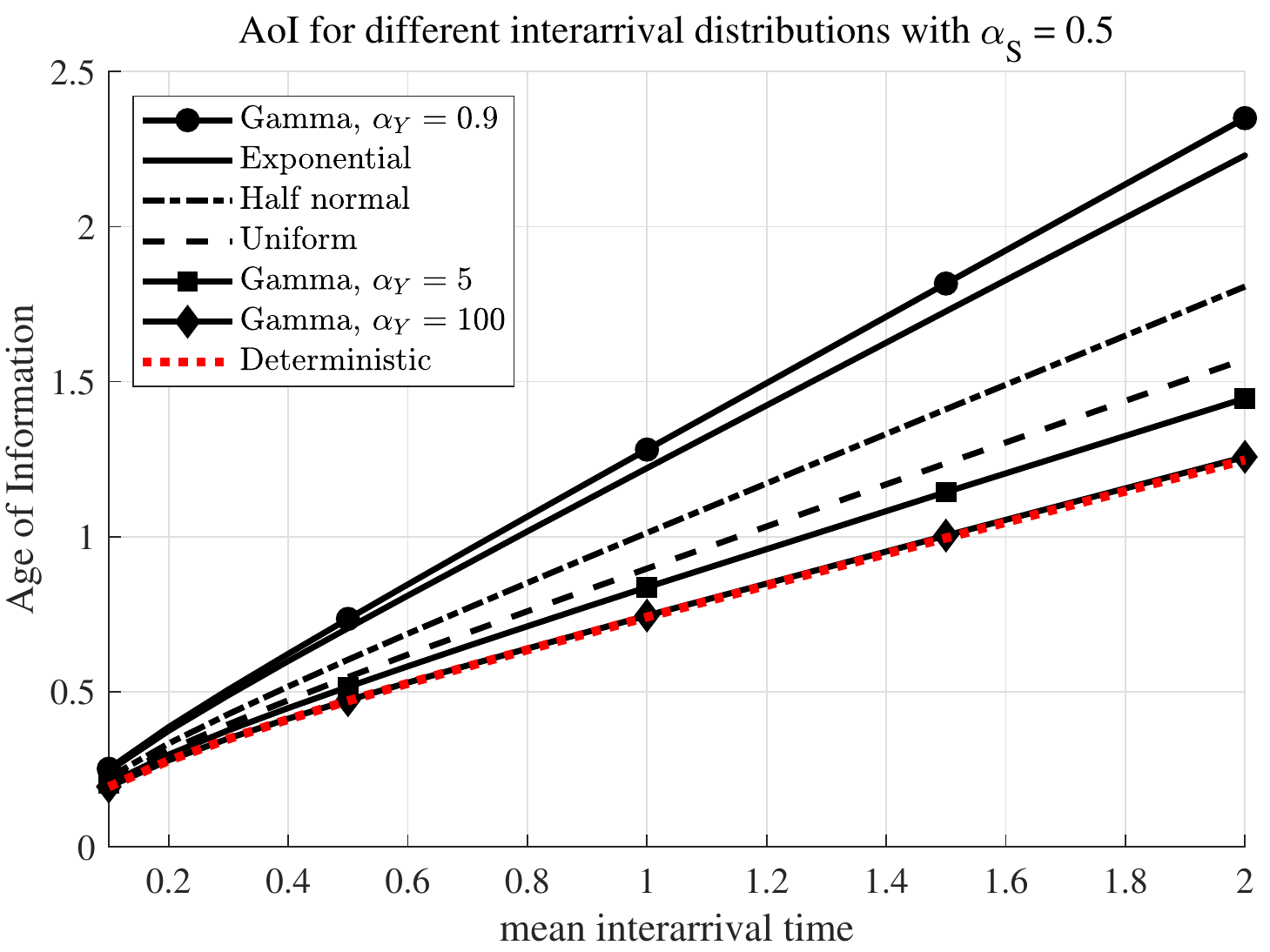}}
	\caption{Average AoI for G/G/1/1 with preemption in service discipline with respect to mean interarrival time for several interarrival distributions.}
	\label{fig:thm-preemption-gg-opt}
\end{figure}

In Fig.~\ref{fig:thm-preemption-gg-opt}(a), we plot the average age for different interarrival time distributions when the service time is gamma with shape parameter $\alpha_S=2$, and rate parameter $\mu=2$. Unlike the case in the blocking discipline, deterministic interarrival times do not result in the minimum age for all mean interarrival values. We observe from Fig.~\ref{fig:thm-preemption-gg-opt}(a) that there is a threshold, above which deterministic interarrivals are optimum and below which exponential interarrivals are optimum when the optimization is over log-concave distributions. On the other hand, in Fig.~\ref{fig:thm-preemption-gg-opt}(b), we plot the same interarrival distributions for a gamma service time with shape parameter $\alpha_S=0.5$, and rate parameter $\mu=2$. Here, irrespective of the mean interarrival time, deterministic interarrivals are always optimum.

\subsection{Age for General Interarrival and Exponential Service Times}

In this section, we consider general interarrival and exponential service times in the preemption in service discipline. We derive the exact age expression that can be written as a summation of two terms, the first of which  depends only on the first and second moments of interarrival times and the second of which depends only on the service rate. In other words, interarrival and service times are decoupled. Average age takes a simple representation as the sum of the age of the arrival process and the mean service time. In addition, it is interesting to see that the time an update spends in service, $\tilde S = S | S < Y$, disappears from the age expression.

\begin{theorem}
\label{thm:preemption-gm}
Consider a G/M/1/1 system with preemption in service, where $Y_n$ are i.i.d. interarrival times with a general distribution and $S_n$ are i.i.d. exponential service times with rate parameter $\mu$. The average age of this system is
\begin{align}
\Delta_\text{G/M}^p =& \frac{E[Y^2]}{2E[Y]} + \frac{1}{\mu}
\label{eqn:thm-preemption-gm}
\end{align}
\end{theorem}

The proof of Theorem~\ref{thm:preemption-gm} is given in Section~\ref{sec:thm-preemption-gm}. The age expression in Theorem~\ref{thm:preemption-gm} is so simple that it only depends on the first moment of the service time, and first and second moments of the interarrival time. When the interarrivals are exponential as well, (\ref{eqn:thm-preemption-gm}) reduces to
\begin{align}
\Delta_\text{M/M}^p =& \frac{1}{\lambda} + \frac{1}{\mu}
\end{align}
which is derived in \cite{Yates17a}.

\begin{corollary}
\label{cor:preemption-gm-opt}
Consider a G/M/1/1 system with preemption in service discipline, where $Y_n$ are i.i.d. interarrival times with a general distribution, and $S_n$ are i.i.d. exponential service times. Given a mean interarrival time constraint, deterministic intearrival times are optimum. In addition, exponential interarrival times result in the worst age when interarrival distribution is log-concave.
\end{corollary}
\begin{Proof}
For a given $E[Y]$, (\ref{eqn:thm-preemption-gm}) is minimized when the second moment, or equivalently, the variance of $Y$ is minimized. Since deterministic variables have zero variance, deterministic interarrival times minimize the average age.

When $Y$ has a log-concave distribution, we have the following relation between the first and second moments of $Y$ from \cite[Proposition 6.A.6]{Marshall07}
\begin{align}
E[Y^2] \leq 2 E^2[Y]
\label{eqn:prf-cor-preemption-gm-op-01}
\end{align}
where the equality is achieved with an exponential distribution. Therefore, we conclude that exponential interarrivals result in the largest possible $\frac{E[Y^2]}{2E[Y]}$ when $Y$ has a log-concave distribution.
\end{Proof}

\begin{figure}
	\centering
	\subfloat[]{\includegraphics[width=.5\textwidth]{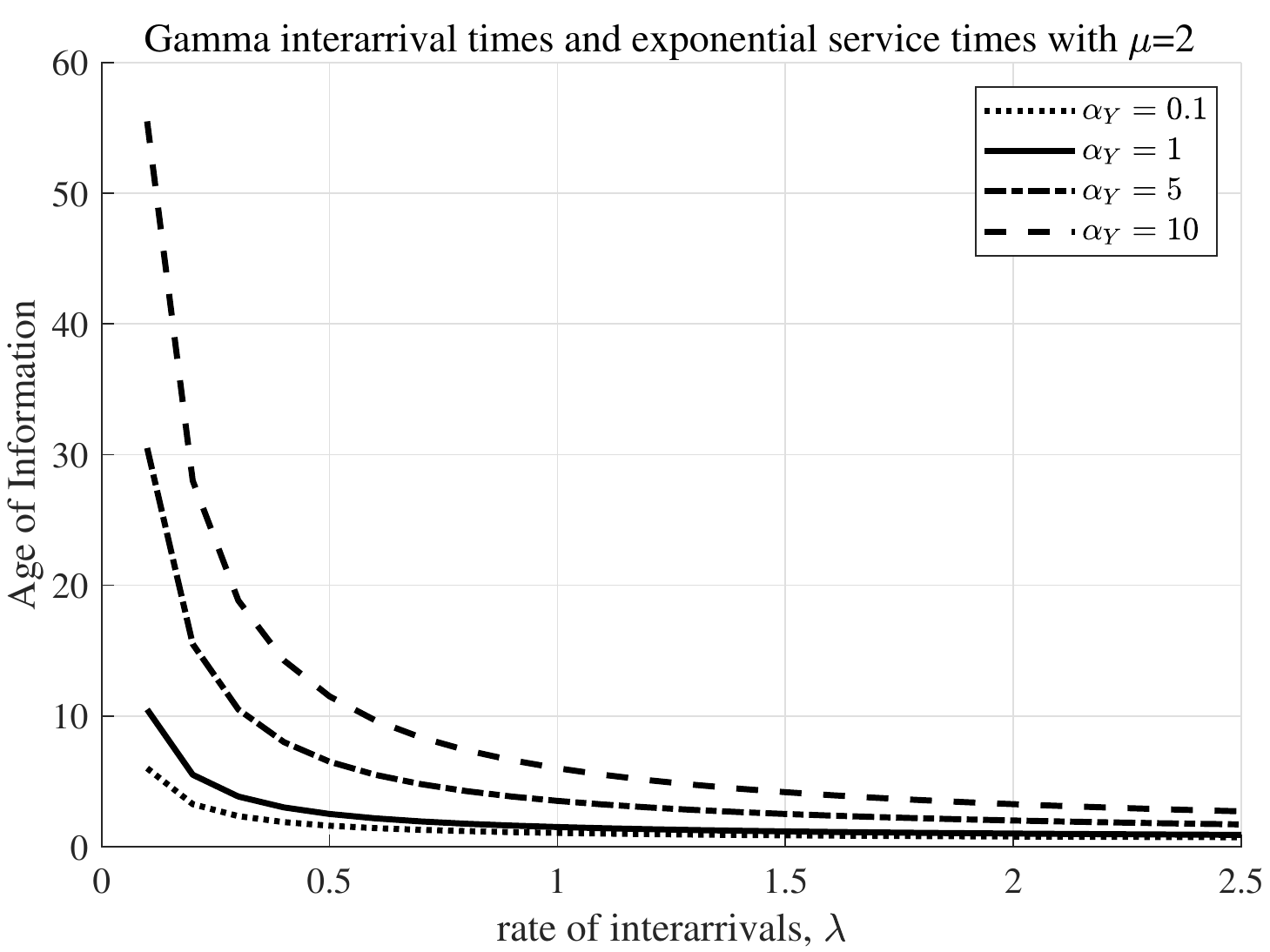}}
	\subfloat[]{\includegraphics[width=.5\textwidth]{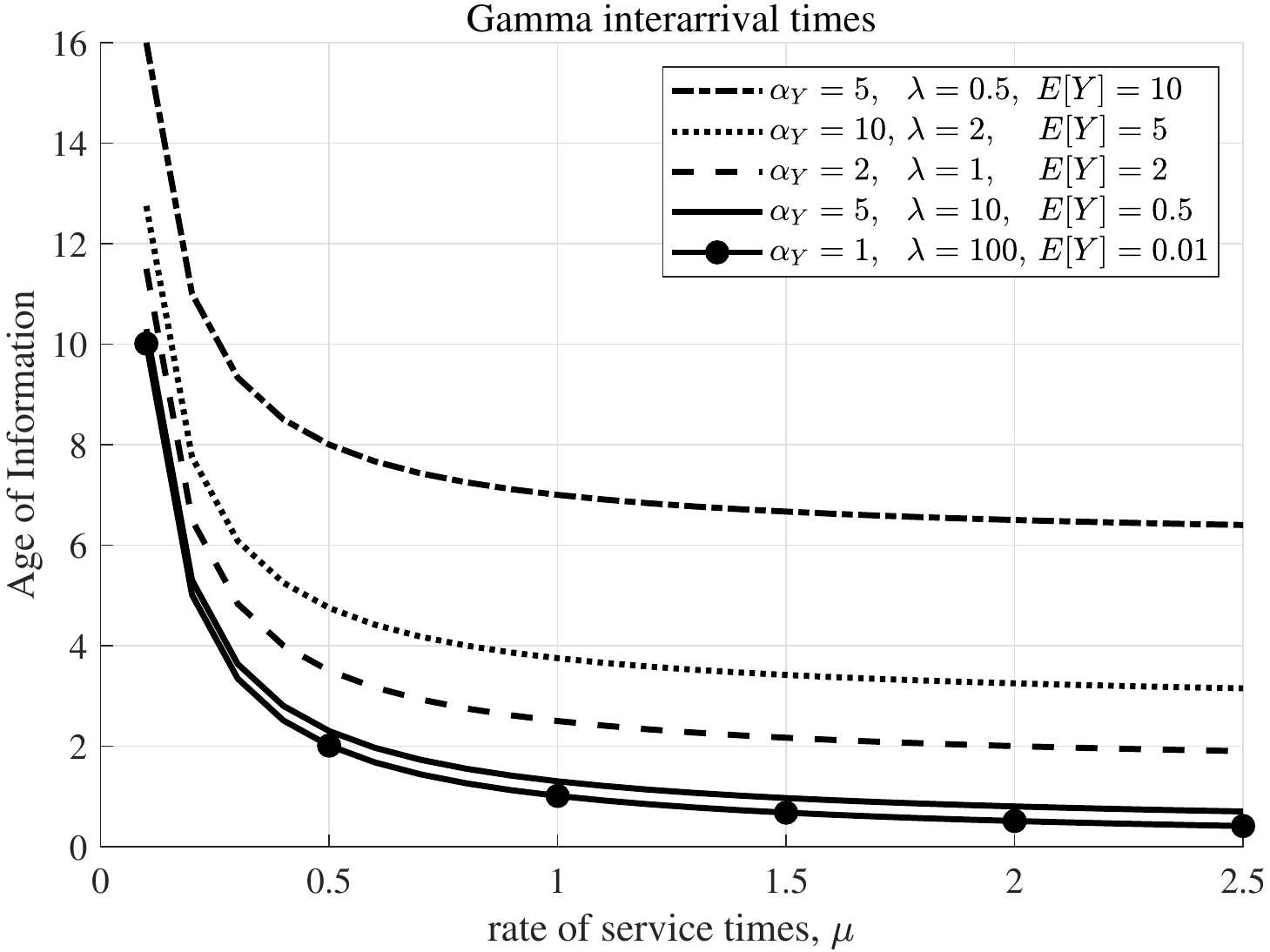}}
	\caption{Average AoI for G/M/1/1 with preemption in service discipline (a) with respect to $\lambda$ when $\mu=2$ and for several $\alpha$, (b) with respect to $\mu$ for several ($\alpha$, $\lambda$) pairs.}
	\label{fig:thm-preemption-gm}
\end{figure}

In Fig.~\ref{fig:thm-preemption-gm}, we consider gamma distributed interarrival times where $\alpha_Y$ is the shape parameter and $\lambda$ is the rate parameter, and exponential service times where $\mu$ is the rate parameter. In Fig.~\ref{fig:thm-preemption-gm}(a), we plot the average age with respect to $\lambda$ when $\alpha_Y$ and $\mu$ are fixed, and in Fig.~\ref{fig:thm-preemption-gm}(b), we plot the average age with respect to $\mu$ when $\alpha_Y$ and $\lambda$ are fixed. We observe that the age decreases with both rate parameters. Fig.~\ref{fig:thm-preemption-gm}(a) shows that smaller $\alpha_Y$ results in a lower age. However, we observe from Fig.~\ref{fig:thm-preemption-gm}(b) that the lowest age is achieved when the mean interarrival time is the smallest.

\subsection{Age for Exponential Interarrival and General Service Times}

In this section, we consider exponential interarrival and general service times, i.e., an M/G/1/1 system, with preemption in service discipline. This system is considered in \cite{Najm17} as well. In Theorem \ref{thm:preemption-mg}, we provide an alternative proof to \cite{Najm17} that determines the age expression in Theorem~\ref{thm:preemption-gg} for an exponential $Y$. Although this is a replication of a previous result, it provides another affirmation that our approach is applicable for many different cases.

\begin{theorem}
\label{thm:preemption-mg}
Consider an M/G/1/1 system with preemption in service, where $Y_n$ are i.i.d. exponential interarrival times with rate parameter $\lambda$ and $S_n$ are i.i.d. service times with a general distribution. The average age of this system is
\begin{align}
\Delta_\text{M/G}^p =& \frac{1}{\lambda E[e^{-\lambda S}]}
\label{eqn:thm-preemption-mg}
\end{align}
\end{theorem}

The proof of Theorem~\ref{thm:preemption-mg} is given in Section~\ref{sec:thm-preemption-mg}.

\section{Conclusions}
Most real world applications require non-exponential interarrival and service time distributions. This paper is an attempt to extend the AoI approach to more practical communication scenarios. We derived exact expressions and upper bounds for the AoI for two service disciplines and for G/G/1/1, G/M/1/1, and M/G/1/1 systems. We observed that the upper bounds are in general close to exact average age expressions. Designing general communication systems with respect to these upper bounds will result in optimized achievable age values.

\section{Appendix}

\subsection{Proof of Theorem~\ref{thm:blocking-gg}}
\label{sec:thm-blocking-gg}

Remember from Section~\ref{sec:model-blocking} that effective interarrival times, $G_n$, can be written as random sums of random numbers; see Fig.~\ref{fig:geometry-blocking}. Although $K$ is not independent of all $Y_j$, it is possible to calculate the expected value of $G$ using Wald's equation \cite[Theorem 3.3.2]{Ross96}, which is stated in Lemma~\ref{lem:wald} below.

\begin{lemma}[Wald's equation {\cite[Theorem 3.3.2]{Ross96}}]
\label{lem:wald}
If $Y_1, Y_2, \dots$ are i.i.d. random variables having finite expectations, and $K$ is a stopping time for $Y_1,Y_2, \dots$ such that $E[K]<\infty$, then
\begin{align}
E\left[ \sum_{k=1}^K Y_k \right] = E[K]E[Y].
\end{align}
\end{lemma}

Using Wald's equation, we have $E[G] = E[K]E[Y]$. Next, we derive an expression for the second moment of the effective interarrival times, $E[G^2]$. Let us first define the indicator function,
\begin{align}
I_k = \left\{
\begin{array}{ll}
1, \quad & \text{if  } k \leq K \\
0, \quad & \text{if  } k > K.
\end{array}
\right.
\label{eqn:prf-blocking-gg-indicator}
\end{align}
Now, we have
\begin{align}
E[G^2] &= E\left[\left( \sum_{k=1}^K Y_k \right)^2 \right] \\
&= E\left[ \left( \sum_{k=1}^\infty Y_k I_k\right)^2 \right]    \\
&= \sum_{k=1}^\infty E[Y_k^2 I_k]  + 2 \sum_{k=2}^\infty\sum_{j=1}^{k-1} E[ Y_k I_k Y_j I_j ].   \label{eqn:prf-blocking-gg-01}
\end{align}
Note that, $I_k =1$ if and only if we have not stopped after successively observing $Y_1, \dots, Y_{k-1}$. Therefore, $I_k$ is determined by $Y_1, \dots, Y_{k-1}$, and is thus independent of $Y_k$. We have $E[Y_k^2 I_k] = E[Y_k^2] E[I_k]$, and $E[ Y_k I_k Y_j I_j ] = E[ Y_k]E [I_k Y_j I_j ]$, for $j<k$. Now, (\ref{eqn:prf-blocking-gg-01}) becomes
\begin{align}
E[ G^2] &= E[Y^2] \sum_{k=1}^\infty E[I_k]  +2E[Y] \sum_{k=2}^\infty\sum_{j=1}^{k-1} E[ I_k Y_j I_j ].
\label{eqn:prf-blocking-gg-02}
\end{align}
First, let us calculate
\begin{align}
\sum_{k=1}^\infty E[I_k] =& \sum_{k=1}^\infty \text{Pr}(K \geq k)   \\
=& \sum_{k=1}^\infty \sum_{j=k}^\infty \text{Pr}(K = j)    \\
=& \sum_{j=1}^\infty \sum_{k=1}^j \text{Pr}(K = j)   \\
=& \sum_{j=1}^\infty j \text{Pr}(K = j)   \\
=& E[K].
\label{eqn:prf-blocking-gg-002}
\end{align}
Next, let us calculate
\begin{align}
\sum_{j=1}^{k-1} E [I_k Y_j I_j]  &= \sum_{j=1}^{k-1}E [Y_j | I_k = 1]\text{Pr}(I_k = 1)   \\
&= E \left[\sum_{j=1}^{k-1}Y_j \, \rule[-12pt]{.45 pt}{28pt} \, I_k = 1\right]\text{Pr}(I_k = 1) \\
&= E \left[\sum_{j=1}^{k-1}Y_j \, \rule[-12pt]{.45 pt}{28pt} \, \sum_{j=1}^{k-1}Y_j < S\right]\text{Pr}\left(\sum_{j=1}^{k-1}Y_j < S\right)
\label{eqn:prf-blocking-gg-03}
\end{align}
where we used the fact that $I_j = 1$ for $j < k$ and given $I_k = 1$.  Note that the condition $I_k=1$ and $\sum_{j=1}^{k-1}Y_j < S$ are equivalent for blocking service discipline. Next, let us denote $A_{k-1} = \sum_{j=1}^{k-1}Y_j$. Then, using Bayes' rule, we can calculate for any $k$,
\begin{align}
E \left[A_k | A_{k} < S\right] = &\int_0^\infty y \frac{\text{Pr}(A_k < S | A_k = y)}{\text{Pr}(A_k<S)}f_{A_k}(y) dy \label{eqn:prf-blocking-gg-04} \\
&= \frac{E[A_k\bar F_S(A_k)]}{\text{Pr}(A_k<S)} \label{eqn:prf-blocking-gg-04-2}
\end{align}
where $A_k$ and $S$ are independent and $\text{Pr}(A_k < S | A_k = y) = \text{Pr}(S > y) = \bar F_S(y)$. Now, (\ref{eqn:prf-blocking-gg-02}) becomes
\begin{align}
E\left[ G^2\right] &= E\left[Y^2\right] E[K]  + 2E[Y] \sum_{k=2}^\infty E[A_{k-1}\bar F_S(A_{k-1})] \\
&= E\left[Y^2\right] E[K]  + 2E[Y] \sum_{k=1}^\infty E[A_k\bar F_S(A_k)]
\label{eqn:prf-blocking-gg-05}
\end{align}

Inserting (\ref{eqn:prf-blocking-gg-05}) and $E[G]=E[K]E[Y]$ into (\ref{eqn:age-blocking-gg}), we have
\begin{align}
\Delta_\text{G/G}^b =& \frac{E[Y^2]}{2E[Y]} + \frac{\sum_{k=1}^\infty E[A_k\bar F_S(A_k)]}{E[K]} + E[S]
\label{eqn:prf-blocking-gg-06}
\end{align}
Now, let us write $E[K]$ in terms of $S$ and $Y$ as
\begin{align}
E[K] = &  \sum_{k=1}^\infty  \text{Pr}(S > A_{k-1}) \\
= & \sum_{k=0}^\infty  \text{Pr}(S > A_{k}) \\
= & 1 + \sum_{k=1}^\infty  E[\bar F_S(A_k)]
\label{eqn:prf-blocking-gg-07}
\end{align}
Finally, inserting (\ref{eqn:prf-blocking-gg-07}) into (\ref{eqn:prf-blocking-gg-06}), we have (\ref{eqn:thm-blocking-gg}).

\subsection{Proof of Lemma~\ref{lem:blocking-gm-geometric}}
\label{sec:lem-blocking-gm-geometric}

Let us start with $\text{Pr} (K=k)$ in (\ref{eqn:lem-blocking-gm-geometric})  as,
\begin{align}
\text{Pr}(K=k) =& \text{Pr}\left( \sum_{j=1}^{k-1} Y_j \leq S < \sum_{j=1}^k Y_j \right)   \\
=& E\left[\bar F_S\left(\sum_{j=1}^{k-1} Y_j \right) - \bar F_S\left(\sum_{j=1}^{k} Y_j \right)  \right]  \\
=& E \left[ e^{-\mu \left(\sum_{j=1}^{k-1} Y_j \right)}  - e^{-\mu \left(\sum_{j=1}^{k} Y_j \right)}  \right]   \\
=& \left( E\left[  e^{-\mu Y } \right] \right)^{k-1} - \left(E\left[  e^{-\mu Y } \right]\right)^{k}   \\
=&  \left( E\left[  e^{-\mu Y } \right] \right)^{k-1} \left( 1 -   E\left[  e^{-\mu Y } \right] \right),
\end{align}
which shows that $K$ is geometric with $p = 1 -   E\left[  e^{-\mu Y } \right] $.

\subsection{Proof of Theorem~\ref{thm:blocking-gm}}
\label{sec:thm-blocking-gm}

Let us start with calculating the middle term on the right hand side of (\ref{eqn:thm-blocking-gg}) for an exponential $S$. First, note that the denominator is $E[K]$ (see (\ref{eqn:prf-blocking-gg-06}) and (\ref{eqn:prf-blocking-gg-07})). Due to Lemma~\ref{lem:blocking-gm-geometric}, $K$ is geometric with $p = 1 -   E\left[  e^{-\mu Y } \right]$ when $S$ is exponential with rate $\mu$. Therefore, we have $E[K] = \frac{1}{1 -   E\left[  e^{-\mu Y } \right]}$. Next, let us consider the numerator. We have
\begin{align}
\sum_{k=1}^\infty E[ A_k\bar F_S(A_k)] =& \sum_{k=1}^\infty E\left[\left(\sum_{j=1}^{k} Y_j \right)  e^{-\mu \left(\sum_{j=1}^{k} Y_j \right) } \right] \\
 =& \sum_{k=1}^\infty E\left[\left(\sum_{j=1}^{k} Y_j \right) \prod_{j=1}^{k} e^{-\mu  Y_j } \right]
 \label{eqn:prf-thm-blocking-gm-001}\\
 =& \sum_{k=1}^\infty \sum_{j=1}^{k} E\left[Y_j  e^{-\mu  Y_j } \prod_{{j '} \neq j}^{k} e^{-\mu  Y_{j '} }\right] \\
=& \sum_{k=1}^\infty \sum_{j=1}^{k} E\left[ Y_j  e^{-\mu  Y_j } \right] \left( E[ e^{-\mu  Y }] \right)^{k-1} \\
=&  E\left[ Y  e^{-\mu  Y}\right]  \sum_{k=1}^\infty k \left( E[ e^{-\mu  Y } ] \right)^{k-1} \\
=&  \frac{E\left[ Y  e^{-\mu Y} \right]  }{\left(1 - E[ e^{-\mu Y}] \right)^{2} }
\label{eqn:prf-thm-blocking-gm-01}
\end{align}
where in (\ref{eqn:prf-thm-blocking-gm-01}), we used the fact that $E[e^{-\mu Y}] < 1$. Since $e^{-\mu Y} < 1$ for every realization of $Y$ except a single point, $Y=0$, which has probability zero. Finally, inserting (\ref{eqn:prf-thm-blocking-gm-01}) into (\ref{eqn:thm-blocking-gg}) with $E[K] = \frac{1}{1 -   E\left[  e^{-\mu Y } \right]}$, we have (\ref{eqn:thm-blocking-gm}).

\subsection{Proof of Corollary~\ref{cor:thm-blocking-gm-equiv}}
\label{sec:cor-thm-blocking-gm-equiv}

Let us start with the middle term on the right hand side of (\ref{eqn:thm-blocking-gm}). For an exponential $S$, we have $E[Ye^{-\mu Y}] = E[Y\bar F_S (Y)]$, which using (\ref{eqn:prf-blocking-gg-04})-(\ref{eqn:prf-blocking-gg-04-2}) can be written as
\begin{align}
E[Ye^{-\mu Y}] = E[Y | Y < S] \text{Pr}(Y < S)
\label{eqn:prf-cor-thm-blocking-gm-equiv-01}
\end{align}
For an exponential $S$, we also have $1 - E[e^{-\mu Y}] = \text{Pr}(Y > S)$. Now, the middle term on the right hand side of (\ref{eqn:thm-blocking-gm}) can be written as
\begin{align}
\frac{E[Ye^{-\mu Y}]}{1 - E[e^{-\mu Y}]} &= \frac{ E[Y | Y < S] \text{Pr}(Y < S)}{\text{Pr}(Y > S)}
\label{eqn:prf-cor-thm-blocking-gm-equiv-02}
\end{align}
When $S$ is exponential, $Y$ is nonnegative and $S$ is independent of $Y$, we have
\begin{align}
E[S] &= E[S - Y | S > Y] \\
&= E[S | S > Y] - E[Y | S > Y]
\label{eqn:prf-cor-thm-blocking-gm-equiv-03}
\end{align}
due to the memoryless property of the exponential distribution. By pulling $E[Y | S > Y]$ from (\ref{eqn:prf-cor-thm-blocking-gm-equiv-03}) and inserting it into (\ref{eqn:prf-cor-thm-blocking-gm-equiv-02}), we have
\begin{align}
\frac{E[Ye^{-\mu Y}]}{1 - E[e^{-\mu Y}]} &= \frac{ E[S | S > Y]\text{Pr}(S > Y) - E[S] (1 - \text{Pr}(Y > S))}{\text{Pr}(Y > S)} \\
&=  \frac{ E[S | S > Y] \text{Pr}(S > Y) - E[S]  +E[S]\text{Pr}(Y > S)}{\text{Pr}(Y > S)}
\label{eqn:prf-cor-thm-blocking-gm-equiv-04}
\end{align}
We know that
\begin{align}
E[S] = E[S | S < Y]\text{Pr}(S < Y) + E[S | S > Y]\text{Pr}(S > Y)
\label{eqn:prf-cor-thm-blocking-gm-equiv-05}
\end{align}
Using (\ref{eqn:prf-cor-thm-blocking-gm-equiv-05}), (\ref{eqn:prf-cor-thm-blocking-gm-equiv-04}) becomes
\begin{align}
\frac{E[Ye^{-\mu Y}]}{1 - E[e^{-\mu Y}]} &=  \frac{- E[S | S < Y]\text{Pr}(S < Y)  +E[S]\text{Pr}(Y > S)}{\text{Pr}(Y > S)} \\
&=  - E[S | S < Y] + E[S]
\label{eqn:prf-cor-thm-blocking-gm-equiv-06}
\end{align}
By inserting (\ref{eqn:prf-cor-thm-blocking-gm-equiv-06}) into (\ref{eqn:thm-blocking-gm}) and noting that $E[S] = \frac{1}{\mu}$, we have (\ref{eqn:cor-thm-blocking-gm-equiv}).

\subsection{Proof of Theorem~\ref{thm:blocking-mg}}
\label{sec:thm-blocking-mg}

Let us start with the middle term on the right hand side of (\ref{eqn:thm-blocking-gg}). Using (\ref{eqn:prf-blocking-gg-04})-(\ref{eqn:prf-blocking-gg-04-2}), the numerator can be written as
\begin{align}
\sum_{k=1}^\infty E [A_k \bar F_S (A_{k}) ] = & \sum_{k=1}^\infty E \left[A_k | A_{k} < S\right] {\text{Pr}(A_k<S)} \\
= & \sum_{k=1}^\infty \int_0^\infty y {\text{Pr}(A_k < S | A_k = y)}f_{A_k}(y) dy \\
= & E_S \left[ \int_0^S  \sum_{k=1}^\infty  y f_{A_k}(y) dy \right]
\label{eqn:prf-thm-blocking-mg-01}
\end{align}
Note that $A_k$ is a sum of $k$ exponentials, which has an Erlang distribution. Using the density of Erlang distribution with rate parameter $\lambda$ and shape parameter $k$, we have
\begin{align}
\sum_{k=1}^\infty y f_{A_k}(y) =& \sum_{k=1}^\infty y \frac{\lambda^{k}y^{k-1}e^{-\lambda y}}{(k-1)!} \\
=& \lambda y e^{-\lambda y} \sum_{k=1}^\infty  \frac{\lambda^{k-1}y^{k-1}}{(k-1)!} \\
=& \lambda y e^{-\lambda y} \sum_{k=0}^\infty  \frac{(\lambda y)^{k}}{k!} \\
=& \lambda y e^{-\lambda y} e^{\lambda y} \\
=& \lambda y
\end{align}
where we used the power series expansion of the exponential function $e^x = \sum_{k=0}^\infty  \frac{x^{k}}{k!}$. Now, (\ref{eqn:prf-thm-blocking-mg-01}) becomes
\begin{align}
\sum_{k=1}^\infty E [A_k \bar F_S (A_{k}) ] = & E_S \left[ \int_0^S  \lambda y dy \right] \\
= & \frac{\lambda E_S [S^2]}{2}
\label{eqn:prf-thm-blocking-mg-02}
\end{align}
Finally, for an exponential $Y$, we know that $G = Y + S$. Using the fact that $E[G] = E[K]E[Y]$, we obtain $E[K] = \frac{E[Y]+E[S]}{E[Y]}$. Inserting $E[K]$ and (\ref{eqn:prf-thm-blocking-mg-02}) into (\ref{eqn:thm-blocking-gg}), we have (\ref{eqn:thm-blocking-mg}).

\subsection{Proof of Theorem~\ref{thm:blocking-lcg}}
\label{sec:thm-blocking-lcg}

We start similar to the proof of Theorem~\ref{thm:blocking-mg}, and aim to prove the following upper bound for $\sum_{k=1}^\infty f_{A_k}(y)$ when interarrival times have log-concave distribution and $u(y)$ is the unit step function:
\begin{align}
\sum_{k=1}^\infty f_{A_k}(y) \leq \frac{u(y)}{E[Y]}
\label{eqn:prf-thm-blocking-lcg-01}
\end{align}
If (\ref{eqn:prf-thm-blocking-lcg-01}) holds, (\ref{eqn:prf-thm-blocking-mg-01}) is upper bounded by $\frac{E[S^2]}{2E[Y]}$, and we have (\ref{eqn:thm-blocking-lcg}). We prove (\ref{eqn:prf-thm-blocking-lcg-01}) by taking the Laplace transform of both sides. Since Laplace transform is a linear operator with non-negative coefficients, the direction of the inequality is preserved. Let us denote the Laplace transforms of $f_Y(y)$ and $f_{A_k}(y)$ as $\phi_Y(\omega)$ and $\phi_{A_k}(\omega)$, respectively. Since $f_{A_k}(y)$ is the $k$-fold convolution of $f_Y(y)$, we have
\begin{align}
\phi_{A_k}(\omega) = \left( \phi_Y(\omega)\right)^k
\end{align}
Note that the Laplace transform of the unit step function is $\frac{1}{\omega}$. Then, by taking the Laplace transform of both sides of (\ref{eqn:prf-thm-blocking-lcg-01}), we have
\begin{align}
\sum_{k=1}^\infty \left( \phi_Y(\omega)\right)^k \leq \frac{1}{E[Y] \omega }
\label{eqn:prf-thm-blocking-lcg-02}
\end{align}
Note that $\phi_Y(\omega) = E[e^{-\omega Y}]$, and since $\omega \geq 0$ for non-negative functions, we have $\phi_Y(\omega) < 1$. Then, (\ref{eqn:prf-thm-blocking-lcg-02}) becomes
\begin{align}
\frac{\phi_Y(\omega)}{1 - \phi_Y(\omega)} \leq& \frac{1}{E[Y] \omega } \\
\phi_Y(\omega) \leq& \frac{1}{1 + E[Y] \omega}
\label{eqn:prf-thm-blocking-lcg-03}
\end{align}
In order to show (\ref{eqn:prf-thm-blocking-lcg-03}), let us use the fact that $Y$ has log-concave distribution and therefore has NBUE property (see Section~\ref{sec:ordering}). We have the following relation for the cumulative distribution function of $Y$, $F_Y(y)$, from \cite[page 174]{Marshall07}
\begin{align}
E[Y] F_Y(y) \leq y - \int_0^y F_Y(t) dt
\label{eqn:prf-thm-blocking-lcg-04}
\end{align}
By taking Laplace transform of both sides, we have
\begin{align}
E[Y] \frac{\phi_Y(\omega)}{\omega} \leq \frac{1}{\omega^2} - \frac{\phi_Y(\omega)}{\omega^2} \\
\phi_Y(\omega) \leq  \frac{1}{1 + E[Y]\omega}
\label{eqn:prf-thm-blocking-lcg-05}
\end{align}
which is exactly (\ref{eqn:prf-thm-blocking-lcg-03}). Finally, taking the inverse Laplace transform of (\ref{eqn:prf-thm-blocking-lcg-02}) completes the proof.

\subsection{Proof of Theorem~\ref{thm:blocking-dn}}
\label{sec:thm-blocking-dn}

In this proof,  we need several definitions and results from total positivity theory \cite{Marshall07} and stochastic ordering \cite{Shaked07}, which are summarized in Section~\ref{sec:ordering}. First, we show in Lemma~\ref{lem:log-concave} below that $G = \sum_{k=1}^K Y_k$ has a log-concave density when interarrival time distribution is log-concave.

\begin{lemma}
Consider a non-negative valued log-concave random variable $S$ that is independent of i.i.d. non-negative valued log-concave random variables $Y_k$, $k=1, \dots, K$ such that $G = \sum_{k=1}^K Y_k$ with ${\rm Pr}(K=k) = {\rm Pr}\left(\sum_{j=1}^{k-1} Y_j \leq S < \sum_{j=1}^k Y_j\right)$. Then, $G$ has a log-concave density as well.
\label{lem:log-concave}
\end{lemma}
\begin{Proof}
First, let us define $g(y,k)$ as the $k$-fold convolution of the probability density function of $Y$, $f_Y(y)$,
\begin{align}
g(y, k) =  f_{Y_1} * f_{Y_2} * \cdots * f_{Y_k} (y)
\end{align}
and $h(k, s)$ as the conditional probability of $K$ given $S=s$,
\begin{align}
h(k, s) = {\rm Pr}\left(\sum_{j=1}^{k-1} Y_j \leq s < \sum_{j=1}^k Y_j\right)
\end{align}
where $S$ is independent of all $Y_j$. Now, we can write the density of $G$ as,
\begin{align}
f_G(y) =& \sum_{k=1}^\infty  f_{Y_1} * f_{Y_2} * \cdots * f_{Y_k} (y) {\rm Pr}(K=k)\\
=& \int_0^\infty \left( \sum_{k=1}^\infty g(y, k)h(k, s) \right)  f_S(s) ds \\
=& \int_0^\infty f_G(y,s) f_S(s) ds
\end{align}
where $f_G(y,s) = \sum_{k=1}^\infty g(y, k) h(k, s)$.

In the remainder of the proof, we use terminology from the theory of total positivity. We refer the reader to Section~\ref{sec:ordering} for a summary of the concepts that follow. For a probability density function that is defined on the non-negative real line, log-concavity, $PF_2$, and $TP_2$ properties imply one another \cite[Proposition 21.B.8]{Marshall07}. Since $f_Y(y)$ and $f_S(s)$ are log-concave due to our hypothesis, they are also $PF_2$ and $TP_2$.

When $f_Y(y)$ is $TP_2$, we conclude that $g(y, k)$ is $TP_2$ from \cite[Theorem 1]{Karlin60} and $h(k, s)$ is $TP_2$ from \cite[Theorem 2]{Karlin60}. Next, when both $g(y, k)$ and $h(k, s)$ are $TP_2$, we conclude that  $f_G(y,s)$ is $TP_2$ from \cite[Lemma 2]{Karlin60}. Finally, when both $f_G(y,s)$ and $f_S(s)$ are $TP_2$, we invoke \cite[Lemma 2]{Karlin60} again and conclude that $f_G(y)$ is $TP_2$ as well. Since $f_G(y)$ is a probability density function that is defined on the non-negative real line, $TP_2$ property implies that it is log-concave as well.
\end{Proof}

Next, we write the waiting time as $W = G - S | G > S$. For a given $S=s$, we note that the random variable $W|S=s$ follows the residual life distribution of $G$, using Definition~\ref{def:rld} in Section~\ref{sec:ordering}.  Therefore, when we take the expected value of $W|S=s$ over $G$, we obtain the mean residual life function of $G$, which is given as
\begin{align}
m_G(s) &= E_G[W|S=s] \\
&= E_G[G-s|G>s]
\end{align}
Since $G$ is log-concave, we know from Section~\ref{sec:ordering} that $m_G(s)$ is a non-increasing function of $s$. Next, we show that  $\text{Cov}(W,S) \leq 0$.
\begin{align}
\text{Cov}(W,S) &= E[WS]  - E[W]E[S] \\
&= E_S \left[ E_G [W|S] S \right]  - E[W]E[S] \\
&= E_S [m_G(S)S] - E[W]E[S] \\
&\leq E_S [m_G(S)]E[S] - E[W]E[S] \label{eqn:sec-thm-blocking-dn-01} \\
& = 0
\end{align}
where (\ref{eqn:sec-thm-blocking-dn-01}) follows from the fact that $m_G(s)$ is non-increasing. Let us consider a random variable $\bar S$ that is i.i.d. with $S$ and independent of $W$. Using the fact that $\text{Cov}(W,S) \leq 0$, one can show that $E[(W+S)^2] \leq E[(W+\bar S)^2]$. Now, (\ref{eqn:age-blocking-gg}) can be upper bounded as
\begin{align}
\Delta_\text{G/G}^b \leq & \frac{E[(W+\bar S)^2]}{2E[W+\bar S]} + E[\bar S].
\label{eqn:prf-thm-blocking-dn-01}
\end{align}
where $W$ and $\bar S$ are independent. Our next goal is to show that the expression in (\ref{eqn:prf-thm-blocking-dn-01}) is less than equal to
\begin{align}
\Delta_\text{M/G}^b =&  \frac{E\left[ (Y^e+ \bar S)^2\right]}{2E[Y^e+ \bar S]} + E[\bar S]
\label{eqn:prf-thm-blocking-dn-age-ub-MG}	
\end{align}
which is the age for M/G/1/1 queues.

We first use the fact that $W|S=s$ follows the residual life distribution of $G$. From \cite[page 182]{Marshall07}, we know that the residual life distribution of a log-concave distribution is log-concave as well. Noting also that $S$ is also log-concave, we conclude that $W$ is log-concave from \cite[Lemma 2]{Karlin60}. Now, we use Lemma~\ref{L/cx-exp} in Section~\ref{sec:ordering} to obtain that $W$ is smaller (in the convex order) than the exponential random variable with the same mean, $W^e$, i.e., $W\leq_\text{cx}W^e$. Equivalently using Lemma~\ref{L/hmrl-cx} in Section~\ref{sec:ordering}, we have $W\leq_\text{hmrl}W^e$. Since $W$ is log-concave and therefore has decreasing mean residual life, we have $E[W] \leq E[Y]$, or equivalently, $E[W^e] \leq E[Y^e]$. Utilizing Definition~\ref{def:hmrl} in Section~\ref{sec:ordering} and noting both $W^e$ and $Y^e$ are exponentials, we conclude that $W^e \leq_\text{hmrl} Y^e$. Moreoever, since $\bar S$ is independent of $W$ and $Y^e$, from  Lemma~\ref{L/sumNBUE} in Section~\ref{sec:ordering}, we have $W+ \bar S\leq_\text{hmrl} Y^e+\bar S$. Finally, using Lemma~\ref{L/age-hmrl} in Section~\ref{sec:ordering} with $\phi(A) = A^2$, we have
\begin{align}
\frac{E[(W+ \bar S)^2]}{E[W+\bar S]} \leq \frac{E[(Y^e+\bar S)^2]}{E[Y^e+\bar S]},
\end{align}
which directly implies $\Delta_\text{G/G}^b \leq \Delta_\text{M/G}^b$.

\subsection{Proof of Theorem~\ref{thm:preemption-gg}}
\label{sec:thm-preemption-gg}

Remember from Section~\ref{sec:model-preemption} that the effective interarrival time, $G=\sum_{k=1}^K Y_k$ is a random sum of random numbers, where $K$ is a geometric random variable. From Lemma~\ref{lem:wald}, we have $E[G] = E[K]E[Y]$. Next, we derive an expression for the second moment of the effective interarrival times, $E[G^2]$. Let us first use the indicator function in (\ref{eqn:prf-blocking-gg-indicator}) and the expansion of $E[G^2]$ in (\ref{eqn:prf-blocking-gg-01}). Similar to the case of blocking discipline, here, $I_k$ is independent of $Y_k$ as well, and therefore, we have (\ref{eqn:prf-blocking-gg-02}). Let us consider
\begin{align}
\sum_{j=1}^{k-1} E [I_k Y_j I_j] \! &= \sum_{j=1}^{k-1} \!E [Y_j | I_k = 1]\text{Pr}(I_k=1) \\
&= (k-1) E [Y | Y < S] E[I_k]
\end{align}
where we used the fact that conditions $I_k = 1$ and $Y_j < S$ are equivalent for the preemption in service discipline and for $j <k$. Now, $E[G^2]$ becomes
\begin{align}
E[ G^2] &= E[Y^2] E[K]  +2E[Y] E [Y | Y < S] \sum_{k=2}^\infty (k-1)  E[I_k]
\label{eqn:prf-thm-preemption-gg-001}
\end{align}
where $\sum_{k=1}^\infty E[I_k] = E[K]$ is shown in (\ref{eqn:prf-blocking-gg-002}). Now, let us calculate
\begin{align}
\sum_{k=2}^\infty (k-1)  E[I_k]  = & \sum_{k=1}^\infty (k-1)  \text{Pr}(K \geq k)  \\
= &  \sum_{k=1}^\infty \sum_{j=k}^\infty  (k-1) \text{Pr}(K = j)   \\
= &\sum_{j=1}^\infty \sum_{k=1}^j (k-1) \text{Pr}(K = j)    \\
= & \sum_{j=1}^\infty \text{Pr}(K = j) \sum_{k=1}^j (k-1)   \\
= & \sum_{j=1}^\infty \frac{(j-1)j}{2} \text{Pr}(K = j)   \\
= & \frac{1}{2}E[K(K-1)],
\end{align}
Thus, we have
\begin{align}
\hspace{-6pt}E[ G^2] = & E[Y^2] E[K]  + E[Y] E[Y|Y<S] E[K(K-1)].
\label{eqn:prf-thm-preemption-gg-01}
\end{align}
Next, using Bayes' rule, we calculate
\begin{align}
E [Y | Y < S] &= \int_0^\infty y \frac{\text{Pr}(Y < S | Y = y)}{\text{Pr}(Y<S)}f_Y(y)dy \\
&= \frac{E[Y\bar F_S(Y)]}{1-p}.
\label{eqn:prf-thm-preemption-gg-01-1}
\end{align}
where $\text{Pr}(Y < S | Y = y) = \text{Pr}(S >y) = \bar F_S(y)$. Now, the average age can be written as
\begin{align}
\Delta_\text{G/G}^p = \frac{E[Y^2]}{2E[Y]} + E[Y\bar F_S(Y)]\frac{E[K(K-1)]}{2E[K](1-p)} + E[\tilde{S}].
\label{eqn:prf-thm-preemption-gg-02}
\end{align}
Since $K$ is geometric with $p= 1 - E[\bar F_S(Y)]$, we have
\begin{align}
\frac{E[K(K-1)]}{2E[K]} = \frac{\frac{2-p}{p^2}-\frac{1}{p}}{\frac{2}{p}} = \frac{1-p}{p}
\label{eqn:prf-thm-preemption-gg-03}
\end{align}
Inserting (\ref{eqn:prf-thm-preemption-gg-03}) into (\ref{eqn:prf-thm-preemption-gg-02}), we have (\ref{eqn:thm-preemption-gg}).

\subsection{Proof of Theorem~\ref{thm:preemption-gm}}
\label{sec:thm-preemption-gm}

Let us start by re-writing (\ref{eqn:thm-preemption-gg}) using (\ref{eqn:prf-thm-preemption-gg-01-1}) and the definition of $\tilde S$ as
\begin{align}
\Delta_\text{G/G}^p =& \frac{E[Y^2]}{2E[Y]} + \frac{E[Y | Y < S]\text{Pr}(Y < S)}{\text{Pr}(S < Y)} + E[S | S < Y]
\label{eqn:prf-thm-preemption-gm-01}
\end{align}
We know that
\begin{align}
E[S] = E[S | S < Y]\text{Pr}(S < Y) + E[S | S > Y]\text{Pr}(S > Y)
\label{eqn:prf-thm-preemption-gm-02}
\end{align}
By pulling $E[S | S < Y]$ from (\ref{eqn:prf-thm-preemption-gm-02}) and inserting it into (\ref{eqn:prf-thm-preemption-gm-01}), we have
\begin{align}
\Delta_\text{G/G}^p =& \frac{E[Y^2]}{2E[Y]} + \frac{E[Y | Y < S]\text{Pr}(Y < S) + E[S] - E[S | S > Y]
\text{Pr}(S > Y)}{\text{Pr}(S < Y)} \\
=&  \frac{E[Y^2]}{2E[Y]} + \frac{E[S] - E[S - Y | S > Y]\text{Pr}(S > Y)}{\text{Pr}(S < Y)}
\label{eqn:prf-thm-preemption-gm-03}
\end{align}
When $S$ is exponential, $Y$ is nonnegative and $S$ is independent of $Y$, $E[S] = E[S - Y | S > Y]$ due to the memoryless property of the exponential distribution. Then, (\ref{eqn:prf-thm-preemption-gm-03}) becomes
\begin{align}
\Delta_\text{G/M}^p =&  \frac{E[Y^2]}{2E[Y]} + \frac{E[S]\left( 1 - \text{Pr}(S > Y)\right)}{\text{Pr}(S < Y)} \\
=&  \frac{E[Y^2]}{2E[Y]} + E[S]
\label{eqn:prf-thm-preemption-gm-04}
\end{align}
which gives (\ref{eqn:thm-preemption-gm}) by noting that $E[S] = \frac{1}{\mu}$.

\subsection{Proof of Theorem~\ref{thm:preemption-mg}}
\label{sec:thm-preemption-mg}

Let us start by re-writing (\ref{eqn:thm-preemption-gg}) as
\begin{align}
\Delta_\text{G/G}^p =& \frac{E[Y^2]}{2E[Y]} + \frac{E[Y | Y < S]\text{Pr}(Y < S)}{\text{Pr}(S < Y)} + E[S | S < Y]
\label{eqn:prf-thm-preemption-mg-01}
\end{align}
For an exponential $Y$, we can calculate
\begin{align}
E[Y|Y<S]\text{Pr}(Y < S) =& \int_0^\infty y \text{Pr}(Y < S | Y = y)f_Y(y) dy \\
=& E_S\left[  \int_0^S y \lambda e^{-\lambda y} dy \right] \\
=& \frac{1 - E[e^{-\lambda S}]}{\lambda} - E[Se^{-\lambda S}]
\label{eqn:prf-thm-preemption-mg-02}
\end{align}
We know that $\text{Pr}(S < Y) = \bar F_Y(S) = E[e^{-\lambda S}]$. Finally, $E[S | S < Y]$ can be calculated as
\begin{align}
E[S | S < Y] =& \int_0^\infty s \frac{\text{Pr}(S < Y | S = s)}{\text{Pr}(S<Y)} f_S(s) ds \\
=&  \frac{E[Se^{-\lambda S}]}{E[e^{-\lambda S}]}
\label{eqn:prf-thm-preemption-mg-03}
\end{align}
Inserting (\ref{eqn:prf-thm-preemption-mg-02}) and (\ref{eqn:prf-thm-preemption-mg-03}) into (\ref{eqn:prf-thm-preemption-mg-01}) gives (\ref{eqn:thm-preemption-mg}).

\subsection{Total Positivity and Stochastic Ordering}
\label{sec:ordering}
In order to obtain some of our bounds, we require certain total positivity and stochastic ordering results. In this section, we provide necessary definitions and results from \cite{Marshall07} and \cite{Shaked07}. We start by defining total positivity of order 2.

\begin{definition}{\rm \textbf{\cite[Chapter 21, B.1]{Marshall07} }}
\label{def:TP2}
Let $A$ and $B$ be subsets of the real line. A function $K(\cdot,\cdot)$ defined on $A \times B$ is said to be totally positive of order 2, denoted $TP_2$, if for all $x_1 < x_2$ and $y_1 < y_2$
\begin{align}
K(x_1,y_1)K(x_2,y_2) - K(x_1,y_2)K(x_2,y_1) \geq 0.
\end{align}
\end{definition}

Total positive functions of order 2 are closely related to P\'olya frequency functions of order 2, which are defined next.
\begin{definition}{\rm \textbf{\cite[Chapter 21, B.7]{Marshall07} }}
\label{def:PF2}
A probability density function $f(\cdot)$ is said to be a P\'olya frequency function of order 2 ($PF_2$) if the function
$K(x, y) = f(y - x)$, $-\infty < x,y < \infty$, is $TP_2$.
\end{definition}

The first result in this section is the equivalence of log-concavity, $PF_2$, and $TP_2$ for a probability density function that is defined on the non-negative real line.

\begin{lemma}{\rm \textbf{\cite[Chapter 21, B.8]{Marshall07} }}
\label{lem:PF2}
The function $K(x, y) = f(y - x)$, $-\infty < x, y < \infty$, is $TP_2$ if and only if $f(\cdot)$ is nonnegative and $log f(\cdot)$ is concave on $(-\infty,\infty)$.

Thus, $f(\cdot)$ is log-concave on $(-\infty,\infty)$ if and only if $f(\cdot)$ is $PF_2$.
\end{lemma}

Next, we define the residual life distribution of a random variable. Residual life distribution is a conditional distribution of the remaining life of a random variable given survival up to time.

\begin{definition}{\rm \textbf{\cite[Chapter 1, B.12]{Marshall07} }}
\label{def:rld}
Let $F$ be a distribution function of a random variable $A$ such that $F(0) = 0$. The residual life distribution $F_t$ of $F$ at $t$ is defined for all $t \geq 0$ such that $\bar F(t) > 0$ by
\begin{align}
\bar F_t(t) = \frac{\bar F(x+t)}{\bar F(t)}, \qquad x \geq 0.
\end{align}
\end{definition}

Mean residual life function, $m_A(t)$, is the mean of distribution $F_t(t)$ and models the expected remaining life of a random variable
\begin{align}
m_A(t) & = \int_0^\infty \bar F_t(x) dx \\
&= E\left[ A - t | A > t \right].
\end{align}
\begin{definition}{\rm \textbf{\cite[Chapter 5, D.1]{Marshall07} }}
\label{def/DMRL}
A random variable A with distribution $F$ and finite mean is said to have a decreasing mean residual life ($F$ is DMRL) if the mean, $m_A(t)$, is decreasing in $t \geq 0$.
\end{definition}
\begin{definition}{\rm \textbf{\cite[Chapter 5, E.1]{Marshall07} }}
\label{Def/NBUE}
A nonnegative random variable $A$ with life distribution $F$ is said to be new better than used in expectation (NBUE) if it has a finite mean that is at least as large as the mean residual life length at time $t$, for all $t \geq 0$, i.e.,
\begin{align}
E[A] \geq  E\left[ A - t | A > t \right] = m_A(t), \qquad \forall t.
\end{align}
\end{definition}

We can conclude immediately from these definitions that log-concavity implies DMRL, and DMRL implies NBUE. Random variables that have DMRL and/or NBUE properties can be ordered stochastically with regards to several ordering definitions. In this paper, we need harmonic mean residual life (hmrl) and convex orders that are especially important while comparing a random variable to an exponential random variable with the same mean.
\begin{definition}{\rm \textbf{\cite[Section 2.B.1]{Shaked07} }}
\label{def:hmrl}
Let $A$ and $B$ be two nonnegative random variables with mean residual life functions $m_A(t)$ and $m_B(t)$, respectively, and suppose that the harmonic averages of $m_A(t)$ and $m_B(t)$ are
comparable as follows
\begin{align}
\left[\frac{1}{t} \int_0^t \frac{1}{m_A(u)}\text{d}u \right]^{-1} \leq \left[\frac{1}{t} \int_0^t \frac{1}{m_B(u)}\text{d}u \right]^{-1}, \quad \forall t > 0.
\end{align}
Then, $A$ is said to be smaller than $B$ in the hmrl order, which is denoted as $A \leq_{\text{hmrl}} B$.
\end{definition}
\begin{definition}{\rm \textbf{\cite[Section 3.A.1]{Shaked07} }}
Let $A$ and $B$ be two random variables such that
\begin{align}
E[ \phi(A) ] \leq E[ \phi(B) ], \quad \text{for all convex functions } \phi.
\end{align}
Then, $A$ is said to be smaller than $B$ in the convex order, which is denoted as $A \leq_{\text{cx}} B$.
\end{definition}

The following lemma shows that NBUE random variables are smaller than exponential random variables with the same means with respect to the convex order.

\begin{lemma}{\rm \textbf{\cite[Theorem 3.A.55]{Shaked07} }}
\label{L/cx-exp}
If $A$ is an NBUE random variable with mean $E[A]$, and $A^e$ is an exponential random variable with mean $E[A]$, then
\begin{align}
A \leq_{\text{cx}} A^e.
\end{align}
\end{lemma}

The following lemma extends the convex order to hmrl order.

\begin{lemma}{\rm \textbf{\cite[eqn. (2.B.7)]{Shaked07} }}
\label{L/hmrl-cx}
Let $A$ and $B$ be two positive random variables. If $E[A]=E[B]$, then
\begin{align}
A \leq_{\text{hmrl}} B \iff  A \leq_{\text{cx}} B.
\end{align}
\end{lemma}

The next lemma shows the closure of hmrl order under convolution.

\begin{lemma}{\rm \textbf{\cite[Lemma 2.B.5]{Shaked07} }}
\label{L/sumNBUE}
If the two almost surely positive random variables $A$ and $B$ are such that $A \leq_\text{hmrl} B$, and if $Z$ is an NBUE nonnegative random variable independent of $A$ and $B$, then
\begin{align}
A +Z \leq_\text{hmrl} B +Z.
\end{align}
\end{lemma}

Finally, we need the following lemma in order to relate hmrl order to average age.

\begin{lemma}{\rm \textbf{\cite[eqn. (2.B.5)]{Shaked07} }}
\label{L/age-hmrl}
Let $A$ and $B$ be two positive random variables. $A \leq_{\text{hmrl}} B$ if and only if
\begin{align}
\frac{E[\phi(A)]}{E[A]} \leq \frac{E[\phi(B)]}{E[B]}
\end{align}
for all increasing convex functions $\phi : [0, \infty) \rightarrow \mathbb{R}$.
\end{lemma}
%

%
%
%
%
%
%\bibliography{../../../../../texmf/bibtex/bib_folder/AoILibrary}
%\bibliographystyle{unsrt}

\end{document}